\documentclass{JFM-FLM_Au}

\usepackage[factor=500]{microtype}

\hypersetup{
  colorlinks = true,
  urlcolor   = blue,
  citecolor  = blue,
  linkcolor = blue,
  anchorcolor = blue
}

\nonstopmode{}

\newcommand{\cmplxi}{{\ensuremath{\mathrm{i}}}}
\newcommand{\conste}{{\ensuremath{\mathrm{e}}}}

\newcommand{\mat}[1]{{\ensuremath{\mathsfbi{#1}}}}
\newcommand{\DD}{\ensuremath{\mathcal{D}}}
\newcommand{\Lat}{\ensuremath{{La}_t}}

\newcommand{\normsq}[1]{{\ensuremath{{\left\lVert{#1}\right\rVert}^2_E}}}
\newcommand{\sublabel}[1]{{\textit{#1}}}

\lefttitle{A. Xuan and L. Shen}
\righttitle{Journal of Fluid Mechanics}

\title{Resolvent model-based analyses of coherent structures in Langmuir turbulence}

\author{Anqing Xuan\aff{1} \and Lian Shen\aff{1}}
\affiliation{%
\aff{1}Department of Mechanical Engineering and Saint Anthony Falls Laboratory, University of Minnesota, Minneapolis, Minnesota 55455, USA
}%

\corresau{Lian Shen, \email{shen@umn.edu}}

\begin{document}

\maketitle

\begin{abstract}
We present an analysis of the coherent structures in Langmuir turbulence, a state of the ocean surface boundary layer driven by the interactions between water waves and wind-induced shear, via a resolvent framework. Langmuir turbulence is characterised by multiscale vortical structures, notably counter-rotating roll pairs known as Langmuir circulations. While classic linear stability analyses of the Craik--Leibovich equations have revealed key instability mechanisms underlying Langmuir circulations, the vortical rolls characteristic of Langmuir turbulence, the present work incorporates the turbulent mean state and varying eddy viscosity using data from large-eddy simulation (LES) to investigate the turbulence dynamics of fully developed Langmuir turbulence. Scale-dependent resolvent analyses reveal a new formation mechanism of two-dimensional circulating rolls and three-dimensional turbulent coherent vortices through linear amplification of sustained harmonic forcing. Moreover, the integrated energy spectra predicted by the principal resolvent modes in response to broadband harmonic forcing capture the dominant spanwise length scales that are consistent with the LES data. These results demonstrate the feasibility of resolvent analyses in capturing key features of multiscale turbulence--wave interactions in the statistical stationary state of Langmuir turbulence.
\end{abstract}


\section{Introduction}
The ocean surface boundary layer features characteristic circulating rolls known as Langmuir circulations, which are formed by the combined influence of wind-driven currents and surface waves~\citep{leibovich1983,thorpe2004}. These circulations organise into pairs of counter-rotating rolls, which induce alternating converging and diverging regions at the water surface and lead to the accumulation of buoyant materials into visible streaks known as windrows. The regime of boundary layer turbulence dominated by Langmuir circulations is referred to as Langmuir turbulence, featuring coherent structures spanning a range of spatial and temporal scales influenced by surface waves. Langmuir turbulence plays a crucial role in the mixing and transport of momentum, mass and heat in the upper ocean, which influences the mixed layer depth and mediates air--sea interactions~\citep{sullivan2010,dasaro2014a}.

Early theoretical work by~\citet{craik1976} established a foundation for understanding the dynamics of Langmuir circulations. \citet{craik1976} developed a wave--current interaction model, known as the Craik--Leibovich (CL) model, to describe the evolution of ocean currents under the influence of surface gravity waves. This model is based on a multiscale asymptotic analysis, which essentially takes an average over a short time to filter out the oscillating wave motions and retains the slowly varying currents. This asymptotic analysis shows that the wave effect after averaging is represented by a vortex force, defined as the cross product of the wave's Stokes drift and the flow vorticity, which affects the current evolution. Subsequent studies~\citep{craik1977,leibovich1977a} showed that the CL model admits an instability mechanism that gives rise to vortical motions resembling the roll cells of Langmuir circulations, thereby validating its theoretical relevance for studying Langmuir circulations. The CL equations have since been extensively applied in theoretical and computational studies of these flows. In recent decades, CL equations have been adopted for large-eddy simulations (LES) of Langmuir turbulence, enabling the resolution of the complex, nonlinear evolution of wave-forced upper-ocean turbulence structures and the analysis of ocean boundary layer dynamics under realistic conditions~\citep[see e.g.][]{skyllingstad1995,harcourt2008,roekel2012,yang2014,chamecki2019,deng2019}.

To elucidate the dynamics of Langmuir circulations, modal analysis techniques, such as stability analyses, have been employed by researchers for theoretical studies.
These theoretical analyses of the CL model have not only established the model's relevance to Langmuir circulations but have also provided key insights into the characteristics of flow structures and their physical origins. Linear stability analysis has been conducted to investigate the onset of Langmuir circulations by examining the evolution of infinitesimal perturbations governed by the CL equations~\citep{craik1977,leibovich1977a,leibovich1980a,leibovich1981,cox_langmuir_1993,leibovich_threedimensional_1993,phillips2001a}. It was found that instability can arise from a weak mean shear current in the presence of Stokes drift induced by surface waves. Specifically, the vertical vorticity associated with the spanwise disturbances in the shear flow is tilted by Stokes drift, leading to the amplification of streamwise rolls. This mechanism, known as CL-II or CL2 instability, is widely recognised as the primary generation process for Langmuir circulations. These analyses also revealed key features of Langmuir circulation observed in the ocean, including their orientation, spacing, and associated surface convergence. Stability analyses have also been performed for shallow-water conditions~\citep{phillips2014}. 

While valuable insights have been gained from classic linear stability analyses, there are limitations in what these analyses can describe about fully developed Langmuir turbulence. Linear stability and related modal analyses focus on the generation and growth of vortical cells by considering the most unstable eigenmode of the CL-derived operator, but can be less representative of the sustained turbulence state and the multiscale feature of Langmuir turbulence. Most traditional stability analyses consider two-dimensional roll structures with infinite streamwise lengths. Although the stability of three-dimensional structures has also been analysed, it was found that the two-dimensional structures are more unstable than three-dimensional modes in unstratified water~\citep{leibovich1980a,leibovich_threedimensional_1993}. As a result, most analyses have focused on the dynamics of two-dimensional rolls. However, in addition to large-scale cells, field observations and LES have shown that Langmuir turbulence comprises coherent vortices with a wide range of spatial and temporal scales~\citep{thorpe2004}. Quasi-streamwise vortices with length scales smaller than those of large-scale Langmuir cells have been identified in LES~\citep{mcwilliams1997,xuan2019a,tsai2023}, particularly near the water surface. These vortical structures resemble the horseshoe vortices found in turbulent shear boundary layers but exhibit distinct length scales, suggesting that they are affected by both the shear current and surface waves. Their inclination in the vertical direction and finite streamwise lengths reflect their full three-dimensionality, deviating from the canonical depiction of Langmuir circulations as elongated, roll-like cells.
The turbulent, multiscale nature of these small-scale vortices likely contributes to the randomness of streaks observed at the ocean surface, as supported by experiments associating surface streaks with small-scale coherent vortices~\citep{melville1998,veron2001}. However, these three-dimensional vortices are not adequately considered in existing stability analyses.
Additionally, classic stability analyses of Langmuir circulations typically assume idealised mean velocity profiles and constant eddy viscosity. These assumptions do not well represent the fully developed turbulence state and may therefore overlook a range of coherent structures in Langmuir turbulence.

In addition to classic stability analyses, resolvent analysis is another powerful modal analysis technique widely used for investigating flow dynamics~\citep{trefethen_hydrodynamic_1993,farrell_stochastic_1993,schmid_stability_2001,jovanovic_componentwise_2005,mckeon_criticallayer_2010,jovanovic_bypass_2021}. Unlike classic stability analysis, which focuses on the asymptotic growth of eigenmodes, resolvent analysis examines a linearised system's response to a harmonic input forcing around a given base flow, providing an input--output view of flow behaviours. By characterising the input--output dynamics, resolvent analysis can identify the dominant linear amplification mechanisms and the associated resolvent modes. Resolvent analysis is not limited to laminar base flows. It can be applied to fully developed, statistically stationary turbulent flows. In such cases, the base flow is the ensemble mean flow, which can be approximated as the time-averaged flow by invoking the assumption of ergodicity. The nonlinear interactions omitted in the linearisation, which are likewise statistically stationary, can be represented by a superposition of harmonic forcing modes that continuously excite the system and sustain the modal response~\citep{mckeon_criticallayer_2010,mckeon_experimental_2013}. This view provides a way to study the self-sustaining mechanisms of turbulent flows. The resolvent analysis has been shown to be effective in revealing dominant coherent structures and their dynamics for various types of turbulent flows, such as canonical wall-bounded turbulence~\citep{moarref_modelbased_2013,illingworth_estimating_2018}, rotating channel flows~\citep{nakashima_reconsideration_2019}, stratified flows~\citep{ahmed_resolvent_2021} and boundary layer separation~\citep{wu_response_2022}. 

In the present study, we seek an improved understanding of Langmuir turbulence via the resolvent framework. We analyse the system's response to harmonic forcing to investigate the dynamics of its fully-developed, self-sustained turbulence state. To achieve this goal, we base our analyses on the average of the turbulent flow extracted from LES, which, in contrast to idealised or laminar base flows, reflects the reduced vertical shear resulting from the highly-efficient turbulent mixing in Langmuir turbulence. Furthermore, we incorporate a vertically varying eddy viscosity, which is also derived from the LES data, into the linearised system. While early linearised analyses typically employed either the molecular viscosity or a constant eddy viscosity to examine flow dynamics, recent studies have shown that using a spatially varying eddy viscosity informed by the turbulent mean flow can improve predictions of flow dynamics, particularly in capturing the energy content of dominant resolvent structures~\citep{morra_relevance_2019,illingworth_estimating_2018,symon_use_2023,saldern_role_2024,zhu_resolvent_2024}. The eddy viscosity may serve as a simple mechanism to partially represent the nonlinear effect of turbulence, which can mimic energy dissipation and transfer by unresolved motions. Linear models augmented with eddy viscosity can thus enhance the ability to predict turbulent flows.

Our resolvent analysis predicts amplification across different length scales corresponding to both full-depth, large-scale Langmuir cells and smaller-scale near-surface turbulence vortices. (In this work, we distinguish between the two types of structures for clarity of nomenclature, but they can be broadly interpreted as Langmuir circulations at different scales considering the influence of surface waves.) The flow structures associated with the principal response mode are consistent with the characteristic vortical motions in Langmuir turbulence.  Furthermore, under harmonic forcing, we find that the linearised system with the turbulence mean state can accurately predict the peak spanwise length scales of vertical velocity fluctuations observed in LES.\@ These results further indicate the potential of resolvent analysis as a useful tool for investigating turbulence--wave interactions and the associated coherent structures in Langmuir turbulence.

To our knowledge, this study is the first reported resolvent analysis-based study of Langmuir turbulence. In addition to effectively capturing coherent structures, resolvent analysis reveals a new mechanism for the emergence of these structures in Langmuir turbulence, where Langmuir vortical structures arise as a forced response to nonlinear interactions intrinsic to turbulent flows. Additionally, this forced mechanism provides a dynamic explanation for the near-surface, small-scale vortices observed in previous simulations and experiments. This new interpretation of Langmuir turbulence dynamics complements the CL-II mechanism revealed by stability analyses, providing a more comprehensive understanding of the dynamics underlying turbulence--wave interactions.

The remainder of this paper is organised as follows: \S{}\,\ref{sec:formulation} introduces the formulation of resolvent analysis for turbulence--wave interactions, along with the setup of LES and the base flow extracted from the simulation data. In \S{}\,\ref{sec:results}, we present and discuss the results of the resolvent analyses, including the model-predicted coherent structures and the energy distribution across scales. Finally, in \S{}\,\ref{sec:conclusions}, we summarise our findings and discuss future work.

\section{Formulation and LES data\label{sec:formulation}}
In the present study, we focus on the canonical scenario of Langmuir turbulence, in which the surface gravity wave and wind-driven shear current are co-aligned. In this section, we first describe the linearised equations used to model the perturbations around the base flow for Langmuir turbulence. We then present the setup of the companion LES that is used to obtain the base flow for the linearised model and to validate the model. Finally, we discuss the base flow configuration employed for setting up the resolvent analysis.

\subsection{Linear model for fluctuations in Langmuir turbulence}
To derive the governing equations for the linearised model describing the turbulence--wave interaction in Langmuir turbulence, we begin with the following wave-averaged momentum equations~\citep{suzuki_understanding_2016}:
\begin{equation}
  \frac{\partial \boldsymbol{v}}{\partial t}+ (\boldsymbol{v}^L\bcdot\bnabla)\boldsymbol{v} = 
  -\bnabla \Phi  + \nu \laplace \boldsymbol{v}
  -\boldsymbol{v} \times (\bnabla \times \boldsymbol{v}^L) - (\boldsymbol{v} \bcdot \bnabla) \boldsymbol{v}^L. \label{eq:wave_averaged}
\end{equation}
Notably, as surface waves carry mean momentum~\citep{stokes1847a,longuet-higgins1953,andrews1978a}, distinguishing between Eulerian and Lagrangian motions in wave--current interaction models is necessary.
In~\eqref{eq:wave_averaged}, $\boldsymbol{v}$ is the Eulerian velocity, $\boldsymbol{U}^s$ is the Stokes drift quantifying the Lagrangian transport of fluid particles by the wave, $\boldsymbol{v}^L=\boldsymbol{v}+\boldsymbol{U}^s$ is the Lagrangian velocity, $\nu$ is the molecular kinematic viscosity, and $\Phi=p+{|\boldsymbol{U}^s|}^2-{|\boldsymbol{v}|}^2$ is the modified pressure, with $p$ being the pressure divided by the water density $\rho$~\citep{holm1996}. For brevity, all pressure variables in this study refer to the pressure normalised by the fluid density. Note that, through vector identities, the above formulation~\eqref{eq:wave_averaged} is mathematically equivalent to the original CL formulation, in which the wave effect appears as a vortex force term $\boldsymbol{U}^s \times (\bnabla\times\boldsymbol{v})$~\citep{leibovich1977a}, as used in the LES in~\S{}\,\ref{sec:simulation_setup}. Here, we present this alternative form because the term $(\boldsymbol{v}^L\bcdot\bnabla)\boldsymbol{v}$ makes explicit that the effective advection velocity in the presence of gravity waves is the Lagrangian velocity $\boldsymbol{v}^L$, which is important for determining the dispersion relation in resolvent analysis.

To obtain the linearised equations, the flow velocity $\boldsymbol{v}$ is decomposed into a mean component and a perturbation, $\boldsymbol{v}=\boldsymbol{U}+ \boldsymbol{u}$, where the mean component $\boldsymbol{U}$ is defined as the time- and plane-averaged flow velocity, i.e.\ $\boldsymbol{U}=\overline{\boldsymbol{v}}$, by assuming statistical stationarity and homogeneity.
Following the approach of~\citet{reynolds_mechanics_1972}, \citet{alamo_linear_2006}, \citet{pujals_note_2009} and~\citet{hwang_linear_2010}, we incorporate a simple vertically varying eddy viscosity into the linearised equations for the perturbations $\boldsymbol{u}$, which are written as
\begin{align}
  &\frac{\partial \boldsymbol{u}}{\partial t} + (\boldsymbol{U}^L \bcdot \bnabla) \boldsymbol{u} 
  + (\boldsymbol{u} \bcdot \bnabla) \boldsymbol{U}^L  +\boldsymbol{u} \times (\bnabla \times \boldsymbol{U}^s)  \nonumber \\
  & = - \nabla p 
  + \nu \bnabla \bcdot \left[ \frac{\nu_T}{\nu} (\bnabla \boldsymbol{u} + \bnabla\boldsymbol{u}^\mathrm{T}) \right]
  + \boldsymbol{d}, \label{eq:perturbation_eq}
\end{align}
\begin{equation}
  \bnabla \bcdot \boldsymbol{u} = 0. \label{eq:perturbation_div}
\end{equation}
Here, the perturbations $\boldsymbol{u}$ are advected by the Lagrangian mean velocity $\boldsymbol{U}^L=\boldsymbol{U} + \boldsymbol{U}^s$, the sum of the Eulerian mean velocity $\boldsymbol{U}$ and the Stokes drift velocity $\boldsymbol{U}^s$. An alternative form for the perturbation equation that explicitly includes the vortex force term associated with the Stokes drift velocity is provided in Appendix~\ref{app:alternative_formulation}. For the case of co-aligned waves and currents in the present study, the mean Eulerian current and Stokes drift velocity are defined as $\boldsymbol{U}={[U(y), 0, 0]}^\mathrm{T}$ and $\boldsymbol{U}^s = {[U^s(y), 0, 0]}^\mathrm{T}$, respectively, which vary in the vertical direction $y$. The term $\boldsymbol{d}$ denotes a forcing term. 

Equation~\eqref{eq:perturbation_eq} may be interpreted as the equation for coherent perturbations within the framework of triple decomposition~\citep{reynolds_mechanics_1972}. The influence of background turbulence on the coherent perturbation, manifested as fluctuating forces, is modelled using the eddy viscosity term via the Boussinesq assumption. The forcing term $\boldsymbol{d}={[d_x, d_y, d_z]}^\mathrm{T}$ accounts for the residual nonlinear effect~\citep[see also][equation 2.6]{reynolds_mechanics_1972}, i.e. $\boldsymbol{d} = -\boldsymbol{u} \bcdot \bnabla \boldsymbol{u} + \overline{\boldsymbol{u} \bcdot \bnabla \boldsymbol{u}}$, where the second term, $\overline{\boldsymbol{u} \bcdot \bnabla \boldsymbol{u}}$, ensures that $\boldsymbol{d}$ has a zero mean, thereby representing only the fluctuating nonlinear forcing. This interpretation has been adopted in previous studies, e.g.~\citet{pujals_note_2009} and \citet{hwang_linear_2010}, and essentially assumes that the effect of background turbulence can be represented using the eddy viscosity term. A more general interpretation considers that the eddy viscosity term partially represents the effect of nonlinear interactions from all scales on the perturbation motions, without attributing it to a specific term~\citep[e.g.][]{pickering_eddy_2019,pickering_optimal_2021}. For example, \citet{hwang_mesolayer_2016} and \citet{symon_energy_2021} found that the eddy viscosity term captures some features of the unresolved energy dissipation~\citep{hwang_mesolayer_2016,symon_energy_2021}. In the present study, we do not seek to define the exact physical origin of $\boldsymbol{d}$. Instead, we treat it as a generic representation of nonlinear interactions that are not explicitly captured by the linearised operator. Previous studies found that its inclusion in the perturbation equations enables simpler forms of forcing (e.g.~white in space and time) to more efficiently predict coherent structures~\citep{symon_use_2023,saldern_role_2024}. Is has also been found that the eddy-viscosity-augmented linear operator improves the descriptions of coherent motions and the predictions of turbulence statistics across various types of turbulent flows~\citep{hwang_mesolayer_2016,illingworth_estimating_2018,morra_relevance_2019,pickering_eddy_2019,symon_use_2023,saldern_role_2024}. 

The governing equations above must be completed with boundary conditions. The full system is driven by a constant shear stress representing the wind shear at the surface $y=0$. At the bottom $y=-H$, where $H$ is given based on the boundary layer depth, a stress-free boundary condition is applied because the shear at the base of the ocean mixed layer is weak~\citep{belcher2012}. In the oceans, the Coriolis force acts to balance the overall momentum within the ocean boundary layer~\citep[see e.g.][]{zikanov2003}. In the present setup, a uniform adverse pressure gradient is introduced to balance the mean momentum. This boundary condition configuration has been used in~\citet{xuan2019a} to isolate and examine the wave effect on the wind-driven oceanic turbulent boundary layer.
Within this framework, the perturbation velocities satisfy the stress-free boundary conditions at both the surface and the bottom:
\begin{subequations}
  \begin{align}
  \frac{\partial{u}}{\partial y}=\frac{\partial w}{\partial y} = 0, \quad & v=0 \quad \text{at } y=0, \label{eq:perturbation_bc_top} \\
  \frac{\partial{u}}{\partial y}=\frac{\partial w}{\partial y} = 0, \quad & v=0 \quad \text{at } y=-H. \label{eq:perturbation_bc_bottom}
  \end{align}
\end{subequations}
The surface shear stress is accounted for by the mean flow, and as a result, the perturbation $\boldsymbol{u}$ satisfies the stress-free condition at the surface.

The pressure term in~\eqref{eq:perturbation_eq} can be eliminated using~\eqref{eq:perturbation_div}, resulting in the following formulation in terms of the vertical perturbed velocity $v$ and vertical vorticity $\omega_y$ to facilitate subsequent dynamic analyses:
\begin{eqnarray}
  \frac{\partial \laplace v}{\partial t} + U^L \frac{\laplace v}{\partial x} - U'' \frac{\partial v}{\partial x}  + {U^s}' \frac{\partial \omega_y}{\partial z} - \nu_T \laplace^2 v - 2 \nu'_T  \frac{\partial \laplace v}{\partial y} \nonumber \\
     -  2 \nu''_T \left( - \frac{\partial^2 v}{\partial x^2} +\frac{\partial^2 v}{\partial y^2} - \frac{\partial^2 v}{\partial z^2} \right) = -\frac{\partial^2 d_x}{\partial x \partial y} -\frac{\partial^2 d_z}{\partial z \partial y} + \frac{\partial^2 d_y}{\partial x^2} + \frac{\partial^2 d_y}{\partial z^2}, \label{eq:lap_v}
\end{eqnarray}
\begin{equation}
  \frac{\partial \omega_y}{\partial t} + U^L \frac{\partial \omega_y}{\partial x} + U'\frac{\partial v}{\partial z} - \nu_T\laplace \omega_y - \nu'_T \frac{\partial \omega_y}{\partial y} = \frac{\partial d_x}{\partial z} - \frac{\partial d_z}{\partial x}, \label{eq:omega_y}
\end{equation}
where $\laplace$ denotes the Laplace operator defined as $\laplace=\partial^2/\partial x^2+\partial^2/\partial y^2+\partial^2/\partial z^2$. The above equations are similar to the classic Orr--Sommerfeld and Squire equations for parallel shear flows but include modifications originating from turbulence--wave interactions. First, the advection velocity of the perturbed motions is the Lagrangian mean velocity $U^L$. Additionally, the term ${U^s}'(\partial \omega_y/\partial z)$ in the $v$ equation is unique to the flows influenced by water waves. This term represents the interaction between the spanwise variations in the vertical vorticity and the vertical gradient of the Stokes drift, which directly affect the dynamics of $v$. This term is thus closely associated with the CL-II instability mechanism, which describes the process of the Stokes drift gradient titling vertical vortices into streamwise vortices. Moreover, we note that this term couples the equations of $v$ and $\omega_y$, in contrast to the classic parallel shear flows without wave effects, where the Squire equation is decoupled from the Orr--Sommerfeld equation.
It should be noted that the formulation of the pressure-eliminated perturbation equations is not unique. For example, \citet{leibovich1977} derived three-dimensional perturbation equations in terms of the vertical and streamwise velocity components. Upon rearranging the terms, the vertical perturbation velocity equation in~\citet{leibovich1977} becomes equivalent to~\eqref{eq:lap_v} in the case of constant eddy viscosity and in the absence of forcing.

For conciseness, we denote the variables $v$ and $\omega_y$ as a state vector defined as $\boldsymbol{\xi} = {[v, \omega_y ]}^\mathrm{T}$.
Given that the fully developed turbulence is statistically homogeneous in the horizontal directions ($x$ and $z$) and stationary in time, we express the state variable $\boldsymbol{\xi}$ and forcing $\boldsymbol{d}$ using a Fourier transform with respect to the homogeneous directions and time:
\begin{equation}
  \boldsymbol{\xi} = \iiint_{-\infty}^{\infty} \hat{\boldsymbol{\xi}}(y; k_x, k_z, \omega) \conste^{\cmplxi(k_x x + k_z z - \omega t)}\, \mathrm{d}k_x \mathrm{d}k_z \mathrm{d}\omega, \label{eq:Fourier_xi}
\end{equation}
\begin{equation}
  \boldsymbol{d} = \iiint_{-\infty}^{\infty} \hat{\boldsymbol{d}}(y; k_x, k_z, \omega) \conste^{\cmplxi(k_x x + k_z z - \omega t)}\, \mathrm{d}k_x \mathrm{d}k_z \mathrm{d}\omega, \label{eq:Fourier_d}
\end{equation}
where $\hat{(\cdot)}$ denotes a quantity in the wavenumber--frequency domain, characterised by the streamwise wavenumber $k_x$, spanwise wavenumber $k_z$ and temporal frequency $\omega$. In other words, we may consider the flow state (or forcing) as a superposition of various Fourier modes $\hat{\boldsymbol{\xi}}$ (or $\hat{\boldsymbol{d}}$), with each mode specified by a triplet $(k_x, k_z, \omega)$. Applying the Fourier transform to the linearised perturbation equations, \eqref{eq:lap_v} and \eqref{eq:omega_y}, yields the equation for a single Fourier mode~\citep[see also][]{jovanovic_componentwise_2005,hwang_linear_2010}:
\begin{equation}
  -(\mathrm{i} \omega \mat{E}+\mat{F}) \hat{\boldsymbol{\xi}} =\mat{B} \hat{\boldsymbol{d}}, \label{eq:perturbation_eqn_operator}
\end{equation}
where
\begin{equation}
  \mat{E} = \left[\begin{array}{cc}
      \hat{\laplace} & 0       \\
      0      & \mat{I}
    \end{array}\right],
\end{equation}
\begin{equation}
  \mat{F}=\left[\begin{array}{cc}
      \mathcal{L}_{OS}   & -\mathrm{i} k_z {U^s}' \\
      -\mathrm{i} k_z U' & \mathcal{L}_{Sq}
    \end{array}\right],
\end{equation}
\begin{equation}
  \mat{B} = \left[\begin{array}{ccc}
      -\cmplxi k_x \DD & -k^2 & -\cmplxi k_z \DD \\
      \cmplxi k_z      & 0    & -\cmplxi k_x    
    \end{array}\right],
\end{equation}
\begin{equation}
  \mathcal{L}_{OS} = -\mathrm{i} k_x U^L \hat{\laplace}+\mathrm{i} k_x U''+v_T {\hat{\laplace}}^2+2 v'_T \DD \hat{\laplace}+v''_T\left(\DD^2+k^2\right), \label{eq:os_op}
\end{equation}
\begin{equation}
  \mathcal{L}_{Sq} = -\mathrm{i} k_x U^L+\nu_T \hat{\laplace}+\nu'_T \DD. \label{eq:squire_op}
\end{equation}
In equations (\ref{eq:perturbation_eqn_operator}--\ref{eq:squire_op}), $\mat{I}$ denotes the identity operator, $k=\sqrt{k_x^2 + k_z^2}$ denotes the magnitude of horizontal wavenumbers $(k_x, k_z)$, $\DD$ denotes the derivative operator with respect to $y$, i.e.\ $\partial_y$, and $\hat{\laplace} = \DD^2 - k^2\mat{I}$ denotes the Laplacian operator in the Fourier space. The boundary conditions for $\hat{\boldsymbol{\xi}}$ that are consistent with the stress-free conditions specified in \eqref{eq:perturbation_bc_top} and \eqref{eq:perturbation_bc_bottom} are given by
\begin{align}
  \hat{v} = \DD^2 \hat{v} = \DD \hat{\omega}_y &= 0 \quad \text{at } y=0, \\
  \hat{v} = \DD^2 \hat{v} = \DD \hat{\omega}_y &= 0 \quad \text{at } y=-H.
\end{align}
Additionally, we note that the state variable $\hat{\boldsymbol{\xi}}$ can be transformed to the Fourier mode of perturbed velocity $\hat{\boldsymbol{u}}={[u, v, w]}^\mathrm{T}$ by
\begin{equation}
  \hat{\boldsymbol{u}}  =\mat{C} \hat{\boldsymbol{\xi}}, \label{eq:perturbation_eqn_to_u}
\end{equation}
\begin{equation}
  \mat{C}= \frac{1}{k^2} \left[\begin{array}{cc}
      \cmplxi k_x \DD & -\cmplxi k_z \\
      k^2             & 0 \\
      \cmplxi k_z \DD & \cmplxi k_x
    \end{array}\right].
\end{equation}

By combining \eqref{eq:perturbation_eqn_operator} and \eqref{eq:perturbation_eqn_to_u}, the perturbed velocity in the Fourier space $\hat{\boldsymbol{u}}$ can be related to the forcing $\hat{\boldsymbol{d}}$ through an input--output formulation:
\begin{equation}
  \hat{\boldsymbol{u}} = \mat{T} \hat{\boldsymbol{d}}, \label{eq:transfer_operator}
\end{equation}
where $\mat{T} = \mat{C}{\left( \mathrm{i} \omega \mat{E} - \mat{F} \right)}^{-1} \mat{B}$ is the transfer operator that connects the input forcing $\hat{\boldsymbol{d}}$ to the velocity response $\hat{\boldsymbol{u}}$. To characterise the input--output dynamics, we can obtain a Schmidt decomposition of the transfer operator $\mat{T}$ as
\begin{equation}
  \mat{T}\hat{\boldsymbol{d}} = \sum_{j=1}^{\infty}{ \sigma_j {\langle \hat{\boldsymbol{d}}, \boldsymbol{\phi}_j}\rangle}_E \boldsymbol{\psi}_j, \label{eq:svd_T}
\end{equation}
where $\left\{ \sigma_j \right\}$ are the singular values in descending orders, and $\left\{ \boldsymbol{\phi}_j \right\}$ and $\left\{\boldsymbol{\psi}_j \right\}$ are known as right-singular and left-singular vectors, respectively. The inner product ${\langle {f},{g} \rangle}_E$ is defined as
\begin{equation}
  {\langle {f},{g} \rangle}_E = \int_{-H}^{0} {g}^* {f}\,\mathrm{d}y,
\end{equation}
where ${(\cdot)}^*$ denotes the complex conjugate transpose, $y=0$ denotes the water surface and $H$ is the depth of the boundary layer. This inner product naturally defines the physically meaningful energy norm as ${\| f \|}_E^2 = {\langle f, f \rangle}_E$.
The right-singular and left-singular vectors, $\left\{ \boldsymbol{\phi}_j \right\}$ and $\left\{\boldsymbol{\psi}_j \right\}$, are orthonormal bases of the input and output spaces, respectively. That is, $\left\{ \boldsymbol{\phi}_j \right\}$ and $\left\{\boldsymbol{\psi}_j \right\}$ satisfy ${\langle {\boldsymbol{\phi}}_i,{\boldsymbol{\phi}}_j \rangle}_E = \delta_{ij}$ and ${\langle {\boldsymbol{\psi}}_i,{\boldsymbol{\psi}}_j \rangle}_E = \delta_{ij}$.
The vectors $\left\{\boldsymbol{\phi}_j \right\}$ span the optimal forcing directions, and the vectors $\left\{ \boldsymbol{\psi}_j \right\}$ span the corresponding velocity response directions. In other words, the relationships between the forcing and velocity response~\eqref{eq:transfer_operator} are represented by pairs of $(\boldsymbol{\phi}_j$, $\boldsymbol{\psi}_j)$ modes, and each pair satisfies
\begin{equation}
  \mat{T} \boldsymbol{\phi}_j = \sigma_j \boldsymbol{\psi}_j,
\end{equation}
where $\sigma_j$, the singular value associated with the $j$-th input--output pair, quantifies the amplification of each forcing mode into its corresponding response mode. In this way, the analyses of the input--output dynamics are transformed into analyses of the resolvent modes and the associated energy amplification.

\begin{figure}
  \centering
  \includegraphics{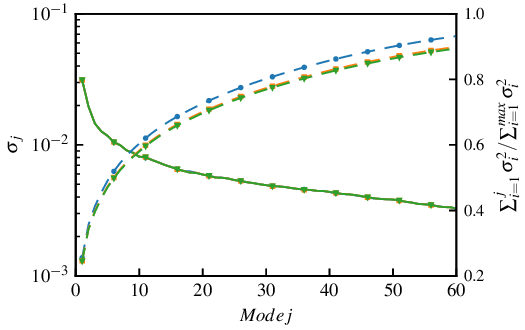}
  \caption{\label{fig:convergence} Convergence of the discretised resolvent system for case $\Lat=0.2$ with $k_x H = 10\pi$, $k_z H=20\pi$ and $\omega= k_x U^L(y=-0.08H) = 459 u_*/H$, as indicated by the singular values, $\sigma_j$ (\full), and the cumulative squared singular values, $\sum_{i=1}^{j} \sigma^2_i / \sum_{i=1}^{max} \sigma^2_i$ (\broken), for different numbers of Chebyshev collocation points: $N=64$ (\fullcirc), $N=128$ (\fullsqr) and $N=256$ (\fulltriangledown). The cumulative squared singular values are plotted against the right vertical axis.}
\end{figure}

In this work, the linearised system (\ref{eq:perturbation_eqn_operator}--\ref{eq:transfer_operator}) is discretised using the rectangular spectral collocation method based on Chebyshev polynomials~\citep{driscoll_rectangular_2016}. This method provides a generalised way to impose boundary conditions with the Chebyshev collocation schemes. More details about the rectangular spectral collocation method are provided in Appendix~\ref{app:collocation}. After the discretised transfer operator is constructed, the decomposition~\eqref{eq:svd_T} is evaluated using the singular value decomposition (SVD) of the matrix representing the operator $\mat{T}$. Through a grid independence study based on the convergence of the singular value spectrum, we decide to use $256$ Chebyshev collocation points of the first kind (see \eqref{eq:collocation_nodes_first} in Appendix~\ref{app:collocation}) in the $y$ direction to impose the differential equations. The grid independence result for an example mode is shown in figure~\ref{fig:convergence}, which compares the results on three Chebyshev grids with $N=64$, $N=128$ and $N=256$. Negligible differences are observed between the results for $N=128$ and $N=256$. This convergence behaviour is observed consistently across the range of wavenumber--frequency triplets considered in the present study, confirming that $N=256$ provides sufficient resolution to capture dominant resolvent modes. 

\subsection{LES setup\label{sec:simulation_setup}}
To enable the linearised model to represent more realistic Langmuir turbulence, we perform LES to extract the base flow state. The simulations also serve as a reference for comparison with the linearised model. The governing equations for the LES utilise the CL equations, which are given as
\begin{equation}
  \frac{\partial \boldsymbol{v}}{\partial t}+ (\boldsymbol{v}\bcdot\bnabla)\boldsymbol{v} = 
    -\bnabla \Pi + \nu \bnabla^2 \boldsymbol{v} - \bnabla\bcdot \tau^{SGS} + \boldsymbol{U}^s\times(\bnabla\times\boldsymbol{v}). \label{eq:CL_LES}
\end{equation}
Here, $\boldsymbol{v}$ denotes the filtered velocity in LES (for brevity, we do not distinguish it notationally from the unfiltered velocity), $\Pi=p + {|\boldsymbol{v}^L|}^2/2-{|\boldsymbol{v}|}^2/2$ is the modified pressure and $\tau^{SGS}$ denotes the subgrid-scale (SGS) stress.
The simulation is set up in a domain with dimensions $L_x \times L_y \times L_z = 8\upi H \times H \times 4\upi H$, where $H$ represents the depth of the ocean surface boundary layer. Periodic boundary conditions are imposed in the horizontal directions $x$ and $z$. In~\citet{deng2019}, a grid size study demonstrated that a domain with dimensions $4 \upi H  \times H \times 8\upi H/3$ (equivalent to $8 \upi h \times 2h \times 16\upi h/3$ in their study, where $h=H/2$ is the half-depth) is sufficiently large to minimise the artificial effects associated with the streamwise and spanwise periodicity. Therefore, the computational domain used in the present LES should be adequate for capturing the largest scale motions. 
The flow is driven by a steady surface wave represented by a constant Stokes drift. Specifically, the Stokes drift is prescribed as that of a deep-water monochromatic wave, $U^s=U^s_0 \exp^{2 k_0 y}$, where $U^s_0$ is the Stokes drift magnitude at the water surface ($y=0$) and $k_0$, the wavenumber of the surface wave, is set to $k_0 H=3.5$. A constant wind-induced shear stress $\tau_w = \rho u_*^2$ is applied at the water surface, with $\rho$ being the water density and $u_*$ being the friction velocity. Langmuir turbulence is characterised by the turbulent Langmuir number, defined as $\Lat=\sqrt{u_*/U^s_0}$, which quantifies the relative importance of the wave forcing to the wind-shear forcing~\citep{mcwilliams1997}. A decreasing $\Lat$ indicates that the wave forcing becomes increasingly dominant. In the present study, we consider two cases, $\Lat=0.2$ and $\Lat=0.3$. Both cases correspond to the Langmuir turbulence regime with strong wave effects~\citep{li2005}. Additionally, we set the friction Reynolds number to $\Rey_\tau=u_* H/\nu=1000$. CL-based LES with moderate Reynolds numbers has been adopted by previous works for investigating turbulence--wave interaction dynamics~\citep{tejada-martinez2007,deng2020}, and has been shown to reproduce the characteristics of Langmuir turbulence observed at higher, practical Reynolds numbers. For example, \citet{deng2020} reported that, for Langmuir circulations in shallow waters, the intensities of Langmuir cells and the magnitudes of Reynolds stress components in most of the water column vary only slightly across a wide range of Reynolds numbers, from $\Rey_\tau=395$ to $\Rey_\tau=10^4$. The observed insensitivity to the Reynolds number is consistent with earlier findings showing the similarity of turbulence statistics across $\Rey_\tau$ from $180$ to $1000$~\citep{tejada-martinez2007,deng2019}, with the LES results in agreement with field measurements. Additionally, for Langmuir turbulence in the open ocean boundary layer, wave-phase-resolved LES at a moderate Reynolds number $\Rey_\tau=2000$ can also predict the vertical turbulence intensities, in good agreement with field measurements~\citep{xuan2020,xuan_effect_2024}. Therefore, the moderate Reynolds number LES here should be able to capture the dominant dynamics of Langmuir turbulence at practical Reynolds numbers and provide a representative dataset for mechanistic studies such as the resolvent analysis in the present work.

The simulations are performed using a hybrid pseudospectral/finite-difference discretisation scheme along with a fractional-step method for time integration. A dynamic Smagorinsky model is used to model the SGS stress~\citep{germano1991,lilly1992}. The computational domain is discretised using a grid of $N_x \times N_y \times N_z = 768 \times 256 \times 1024$, with the grid clustered near the surface and bottom to resolve the boundary layers. The minimum grid spacing in the wall units is $\Delta y^+_{min}=\Delta y_{min} u_*/\nu=0.25$. The numerical method used in these simulations has been validated for LES of Langmuir turbulence in~\citet{deng2019}, and the tests are thus not repeated here.


\subsection{Base flow}
The base flow used in the linearised model is obtained from the mean turbulent velocity profile extracted from the LES result, allowing the resolvent model to more truthfully represent the dynamics of fully developed Langmuir turbulence. Once the mean velocity and Reynolds shear stress profile are obtained, the turbulent eddy viscosity $\nu_t$ is calculated as $\nu_t(y) = -\overline{u'v'}/(\mathrm{d} U^L/\mathrm{d} y)$. Notably, the eddy viscosity here is determined using the gradient of the Lagrangian mean velocity. This approach has been shown to more accurately model turbulence mixing and transport in the ocean surface boundary layer governed by Langmuir turbulence~\citep{mcwilliams2000,yang2015,liu_evaluation_2022}.

\begin{figure}
  \centering
  \includegraphics{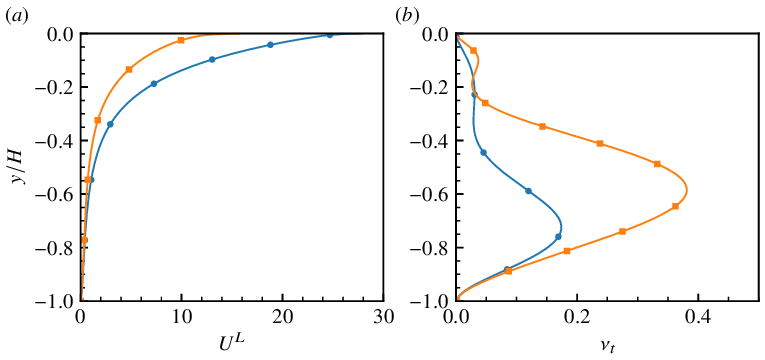}
  \caption{\label{fig:mean_profile}Profiles of (\sublabel{a}) the Lagrangian mean velocity $U^L$ and (\sublabel{b}) the eddy viscosity $\nu_t$ extracted from the simulations of cases with $\Lat=0.2$ (\fullcirc) and $\Lat=0.3$ (\fullsquare).}
\end{figure}

Figures~\ref{fig:mean_profile}(\sublabel{a}) and~\ref{fig:mean_profile}(\sublabel{b}) show the profiles of the mean Lagrangian velocity and eddy viscosity, respectively. Owing to the strong mixing effect of Langmuir turbulence, the Eulerian mean current is nearly negligible throughout most of the boundary layer. As a result, the Lagrangian mean current is predominantly determined by the Stokes drift velocity, which increases exponentially towards the surface ($y=0$).

For the cases considered, the linearised system, described by~\eqref{eq:lap_v} and~\eqref{eq:omega_y} minus the forcing term $\boldsymbol{d}$, is asymptotically stable. In other words, when the turbulent mean velocity and eddy viscosity profiles are considered, the wave--current interaction does not support the long-term growth of infinitesimal perturbations into coherent structures such as Langmuir cells. The asymptotic stability of the system suggests that the traditional stability analysis does not depict the full picture of the dynamics of Langmuir turbulence, for which we present the resolvent analysis below to complement existing studies.

\section{Results\label{sec:results}}
In this section, we present the results obtained from the analyses of the resolvent model for turbulence--wave interactions in Langmuir turbulence. The optimal harmonic responses and the associated forcing and response structures are presented in \S{}\,\ref{sec:modes}.  The spectral behaviour of the coherent structures predicted by the model is discussed in \S{}\,\ref{sec:energy}.

\subsection{Harmonic responses and coherent structures\label{sec:modes}}
For a harmonic forcing with frequency $\omega$, the steady-state response of the linearised system~\eqref{eq:transfer_operator} is also harmonic. The optimal amplification of energy from the forcing to response for a triplet $(k_x, k_z, \omega)$, $G(k_x, k_z, \omega)$, is defined as
\begin{equation}
  G(k_x, k_z, \omega) = \max_{\boldsymbol{d}\neq 0} \frac{{\normsq{\hat{\boldsymbol{u}}}}}{\normsq{\hat{\boldsymbol{d}}}} = \max_{\boldsymbol{d}\neq 0} \frac{{\normsq{\mat{T}\hat{\boldsymbol{d}}}}}{\normsq{\hat{\boldsymbol{d}}}}.
\end{equation}
By definition~\eqref{eq:svd_T}, the amplification is maximised when the input forcing $\hat{\boldsymbol{d}}$ aligns with the principal right-singular mode $\boldsymbol{\phi}_1$ of the decomposition of $\mat{T}$, and the response is $\hat{\boldsymbol{u}}=\sigma_1 \boldsymbol{\psi}_1$. The optimal energy amplification $G$ is thus given by the square of the largest singular value, i.e.\ $G=\sigma_1^2$.

\begin{figure}
  \centering
  \includegraphics{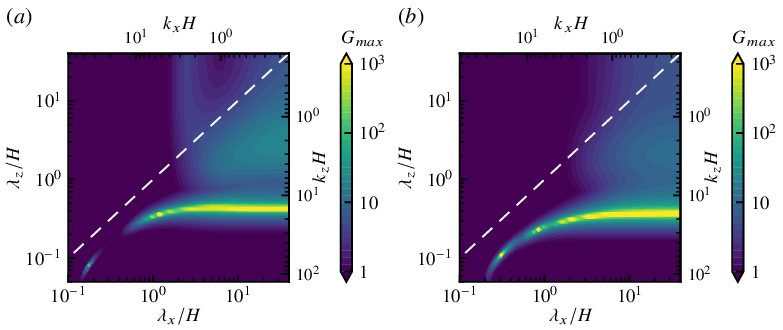}
  \caption{\label{fig:resolvent_Hnorm_2d}Contours of the maximum energy amplification $G_{max}$~\eqref{eq:Gmax} at different streamwise and spanwise wavelengths for cases with (\sublabel{a}) $\Lat=0.2$ and (\sublabel{b}) $\Lat=0.3$. The dashed line indicates $\lambda_x = \lambda_z$.}
\end{figure}

We first consider the maximum energy amplification achieved across all frequencies, which is defined as
\begin{equation}
  G_{max} (k_x, k_z) = \max_{\omega} G(k_x, k_z, \omega). \label{eq:Gmax}
\end{equation}
This maximum amplification is also known as the $H_\infty$ norm of the transfer operator $\mat{T}$~\citep{jovanovic_componentwise_2005}. The computed $G_{max}(k_x, k_z)$ is plotted in figure~\ref{fig:resolvent_Hnorm_2d}, which shows the optimal linear amplification supported by the system across different structure length scales $(\lambda_x, \lambda_z) = (2\upi/k_x, 2\upi/k_z)$. The wavelengths are sampled logarithmically over the ranges of $\lambda_x \in [0.1H, 40H]$ and $\lambda_z \in [0.05H, 40H]$ using a grid of $96 \times 112$ sampling points. For each wavelength pair, the maximum energy amplification is computed by searching over a frequency range corresponding to phase speeds $c \in [0.01 U^L_{max}, U^L_{max}]$, evaluated at $50$ evenly spaced phase speed values, where $U^L_{max}$ denotes the maximum Lagrangian mean velocity, reached at the water surface (see figure~\ref{fig:mean_profile}\sublabel{a}). The corresponding frequency $\omega$ is given by $\omega = c k_x$. For both turbulent Langmuir numbers, $\Lat=0.2$ and $\Lat=0.3$, strong amplification is observed for streamwise elongated structures with $\lambda_x \gg \lambda_z$. In other words, even weak nonlinear forcing inherent to turbulent flows is likely to excite elongated structures. This feature of amplified length scales is consistent with the characteristics of coherent structures in Langmuir turbulence, which are dominated by elongated quasi-streamwise vortices~\citep{mcwilliams1997,teixeira2010,xuan2019a}.

Furthermore, we observe that the amplified structures can be classified into two regimes based on the spanwise wavelengths. Specifically, the amplified structures can be roughly categorised into very large-scale motions with $\lambda_z>H$ and smaller-scale motions with $\lambda_z < H$. For the first regime (very large scales), the maximum amplification $G_{max}$ increases with the streamwise wavelength and $G_{max}$ is found to reach a maximum for streamwise-invariant modes ($k_x=0$, not shown on the logarithmic scale in figure~\ref{fig:resolvent_Hnorm_2d}). This result is also somewhat consistent with previous stability analyses using constant eddy viscosity, which revealed that the streamwise-invariant mode is the most unstable~\citep{leibovich1980a}. In our analysis, this streamwise-invariant mode, corresponding to a static resolvent with $\omega=0$, represents the limiting case of a base flow modification, i.e.\ a two-dimensional mean flow with three velocity components. The strong amplification of the streamwise-invariant mode is also observed in the analysis of turbulent channel flow without the wave forcing~\citep[e.g.][]{hwang_linear_2010}. However, these structures may not clearly manifest in actual flows, likely because nonlinear interactions redistribute energy away from this mode and a persistent nonlinear forcing to close the self-sustaining loop is not evident. Nevertheless, the strong amplification of such limiting modes suggests the presence of streamwise elongated structures with large $\lambda_x/\lambda_z$ aspect ratios. For the second regime consisting of structures with smaller spanwise wavelengths, figure~\ref{fig:resolvent_Hnorm_2d} shows that strong amplification occurs for a range of large streamwise wavelengths with comparable amplification factors, indicating that the linear amplification mechanism is not particularly selective about the streamwise wavelengths for these narrow structures. Moreover, the amplifications in this regime are considerably greater than those of very large-scale structures, suggesting the potential importance of smaller-scale, three-dimensional coherent structures in Langmuir turbulence. 

To further elucidate the amplification dynamics, we next examine the spatial structure of the principal resolvent mode pair $(\boldsymbol{\phi}_1, \boldsymbol{\psi}_1)$, corresponding to the optimal input forcing and the resulting coherent structure supported by the linearised system. Here, we use the modes predicted for case $\Lat=0.2$ as examples for discussion, as the characteristic features of the flow structures are qualitatively similar for the $\Lat=0.3$ case.

\begin{figure}
  \centering
  \includegraphics{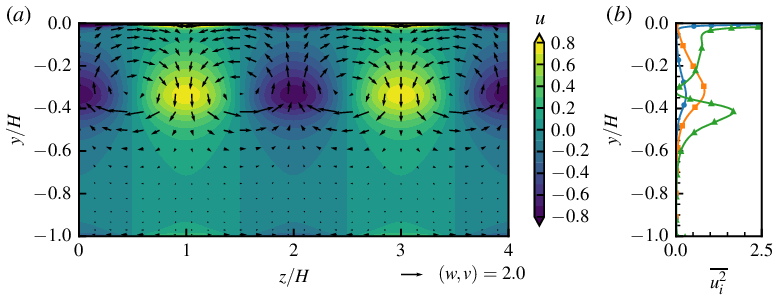}
  \includegraphics{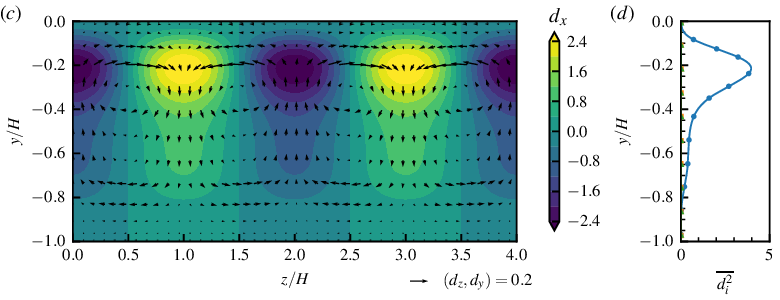}
  \caption{\label{fig:resolvent_structure_2d}Structures of the optimal (\sublabel{a}) velocity response $\boldsymbol{u}=(u,v,w)$ and (\sublabel{c}) input forcing $\boldsymbol{d}=(d_x,d_y,d_z)$ for case $\Lat=0.2$ at $(k_x, k_z, \omega)=(0, 2\upi/H, 0)$. In (\sublabel{a}) and (\sublabel{c}), the contours represent the streamwise component, $u$ and $d_x$, respectively; the vectors represent the cross-stream components, $(w,v)$ and $(d_z, d_y)$, respectively. The vertical variations in the mean squared response velocity and forcing components are plotted in (\sublabel{b}) and (\sublabel{d}), respectively: streamwise component ($\overline{u^2}$ or $\overline{d_x^2}$, \fullcirc), vertical component ($\overline{v^2}$ and $\overline{d_y^2}$, \fullsqr) and spanwise component ($\overline{w^2}$ and $\overline{d_z^2}$, \fulltri).}
\end{figure}

A representative principal mode corresponding to the streamwise-invariant very large-scale motions with $k_x=0$ and $k_z=2\upi/H$ (or $\lambda_x=\infty$ and $\lambda_z=H$) is shown in figure~\ref{fig:resolvent_structure_2d}. For streamwise-invariant motions, it is natural to set the frequency to $\omega=0$, i.e.\ we consider a stationary streamwise-invariant structure. We can observe that the structures consist of pairs of counter-rotating rolls (figure~\ref{fig:resolvent_structure_2d}\sublabel{a}). These streamwise rolls exhibit significant vertical penetration, occupying most of the water depth. This type of structure thus corresponds to large-scale Langmuir cells. In the surface converging zones induced by counter-rotating motions, the streamwise velocity is positive, indicating that the fluid flows faster than the mean flow. Conversely, in the surface diverging zones, the streamwise velocity is lower than the mean velocity. This correspondence between the converging/diverging zones and variations in the streamwise velocity is also consistent with the canonical features of Langmuir cells~\citep{leibovich1983}.
Figure~\ref{fig:resolvent_structure_2d}(\sublabel{b}) shows the mean-square amplitudes of the three velocity components of the principal response mode. The energy of the spanwise and vertical velocities is considerably greater than that of the streamwise velocity, indicating the dominance of streamwise vortical motions. The spanwise velocity nearly vanishes at the depth where the vertical velocity fluctuations reach a maximum, corresponding to the depth of the rotational centre of the streamwise rolls. 

The structure and strength of the corresponding optimal forcing mode are shown in figures~\ref{fig:resolvent_structure_2d}(\sublabel{c}) and~\ref{fig:resolvent_structure_2d}(\sublabel{d}), respectively. The forcing is dominated by the streamwise component. Recall that the forcing (partially) accounts for the nonlinear interactions affecting the perturbation motions, which arise from the self-interaction of the perturbation motions and/or from the interactions with the background turbulence. Therefore, this result shows a mechanism for sustaining the streamwise vortical motions: spanwise variations in streamwise accelerations, as shown in figure~\ref{fig:resolvent_structure_2d}(\sublabel{c}), can excite the system to generate streamwise rolls. This behaviour bears similarity to the CL-II instability mechanism, which states that an initial spanwise perturbation in the streamwise current velocity, in the presence of waves, can grow into Langmuir circulations. The resolvent analysis complements the CL-II mechanism by showing that, in fully developed Langmuir turbulence, large-scale Langmuir cells can be sustained by the variations in the forcing in the streamwise momentum equation. Although investigating the detailed flow processes responsible for such forcing in Langmuir turbulence is beyond the scope of this study, we note that previous studies have shown that the turbulence fluctuations and Reynolds stresses are modulated by large-scale Langmuir cells~\citep{deng2019,deng2020} and manifest spatial correlations with the cells, which suggests that nonlinear interactions in turbulence may indeed produce spatially varying forcing at the wavelength of the amplified mode. Notably, the energy amplification factor between the forcing and response is large (figure~\ref{fig:resolvent_Hnorm_2d}); therefore, a relatively weak nonlinear forcing is capable of generating these circulating motions.

We next examine a representative three-dimensional mode predicted by the model. As shown in figure~\ref{fig:resolvent_Hnorm_2d}, the model predicts strong amplification of structures with small spanwise wavelengths and finite streamwise wavelengths. However, such turbulent coherent structures are often overlooked in classic stability analyses of Langmuir circulations, which typically focus on streamwise-invariant modes. For our analysis, we select the wavenumber and frequency triplet $(k_x, k_z, \omega)$ based on the energetic structure scales observed in the LES of Langmuir turbulence at $\Lat=0.2$. Assuming that the coherent structures are convected by the local mean Lagrangian velocity, the velocity response takes the form of a travelling wave with frequency $\omega = c k_x$ (see equation~\ref{eq:Fourier_xi}), where $c$ is the phase speed and is set to $c=U^L$. Here, we set $c=U^L(y=-0.12H)$, the depth at which the vertical turbulent kinetic energy reaches maximum. Then the streamwise and spanwise wavenumbers $k_x$ and $k_z$ are chosen to be the most energetic wavenumbers in the energy spectra computed at this depth.

\begin{figure}
  \begin{center}
  \includegraphics{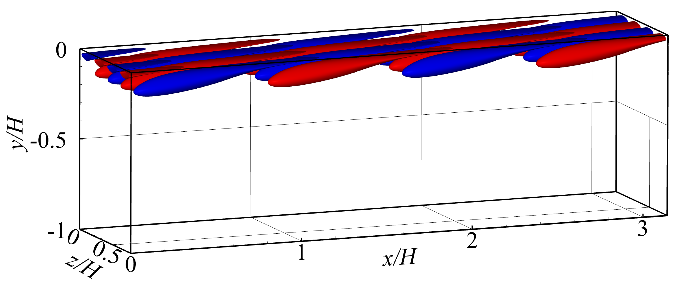}
  \end{center}
  \caption{\label{fig:resolvent_structure_3d}Vortex structures of the principal response mode for case $\Lat=0.2$ with $k_x H=4$, $k_z H = 16.5$ and $\omega=k_x U^L(y=-0.12H)=44.9 u_*/H$. The vortex structures are elucidated using the iso-surfaces of the $Q$-criterion ($10\%$ of the maximum value), with red and blue indicating positive and negative streamwise vorticity, respectively.}
\end{figure}

Figure~\ref{fig:resolvent_structure_3d} shows the flow structures of the selected three-dimensional mode. Using the Q-criterion, the coherent vortical structures revealed are quasi-streamwise vortices, which are most prominent near the water surface. These vortices exhibit alternating rotation directions and are inclined with their near-surface ends pointing in the downstream direction. These structures closely resemble the statistically dominant vortical structures extracted from numerical simulations of Langmuir turbulence~\citep{mcwilliams1997,xuan2019a,tsai2023}.

\begin{figure}
  \includegraphics{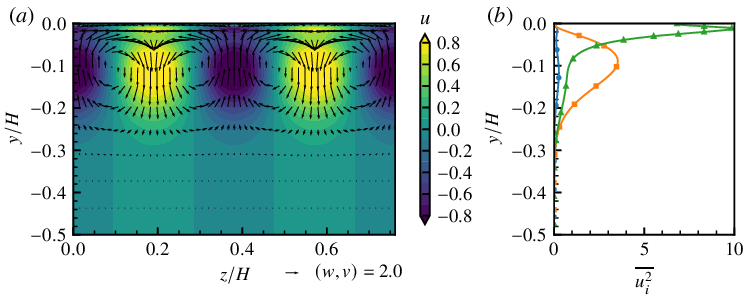}
  \includegraphics{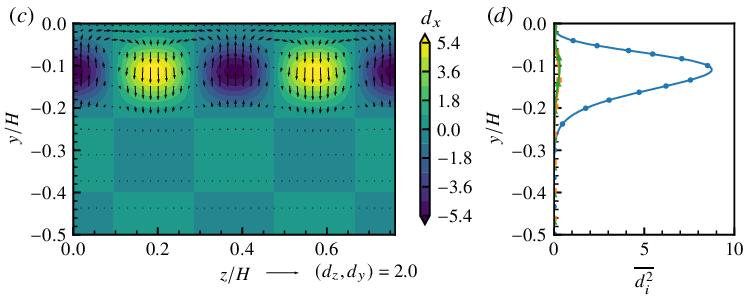}
  \caption{\label{fig:resolvent_structure_3d_velocity}Cross-plane structures of the optimal (\sublabel{a}) response $\boldsymbol{u}=(u, v, w)$ and (\sublabel{c}) input forcing $\boldsymbol{d}=(d_x, d_y, d_z)$ shown in figure~\ref{fig:resolvent_structure_3d}, plotted at $x=(\upi/12)H$. In (\sublabel{a}) and (\sublabel{c}), the contours represent the streamwise component, $u$ and $d_x$, respectively; the vectors represent the cross-stream components, $(w,v)$ and $(d_z, d_y)$, respectively. The vertical variations in the plane-averaged squared response velocity and forcing components are plotted in (\sublabel{b}) and (\sublabel{d}), respectively: streamwise component ($\overline{u^2}$ or $\overline{d_x^2}$, \fullcirc), vertical component ($\overline{v^2}$ and $\overline{d_y^2}$, \fullsqr) and spanwise component ($\overline{w^2}$ and $\overline{d_z^2}$, \fulltri).}
\end{figure}

Figure~\ref{fig:resolvent_structure_3d_velocity}(\sublabel{a}) plots the velocity components on a vertical cross-plane, clearly showing the vortex-induced counter-rotating motions. In the converging zone between two counter-rotating vortices, a downward jet is formed, accelerating the downstream flow away from the surface. In the diverging zone, an upward motion that brings low-momentum fluid towards the surface is observed. Figure~\ref{fig:resolvent_structure_3d_velocity}(\sublabel{b}) shows the energy of the three velocity components. Compared with the large-scale Langmuir cells (figure~\ref{fig:resolvent_structure_2d}\sublabel{b}), the energy of these three-dimensional vortices is more concentrated near the surface. The vertical and spanwise velocity components continue to dominate the energy content of the coherent quasi-streamwise vortices. The vertical velocity fluctuations peak near $y=-0.12H$, coinciding with the depth where $\overline{U}^L=\omega/k_x$. As the free surface is approached, owing to the kinematic constraint imposed by the free-surface boundary conditions, the vertical velocity fluctuations diminish, whereas the spanwise velocity fluctuations increase rapidly. The streamwise velocity fluctuations remain negligible in this near-surface region. The relative strength among the three velocity components for the turbulent coherent structures is consistent with the characteristics of Langmuir turbulence, where the streamwise velocity fluctuations are suppressed by the enhanced mixing induced by strong vertical velocity fluctuations~\citep{li2005}. This behaviour differs from that in classic wall-bounded shear turbulence, where the streamwise component dominates the kinetic energy, indicating that the wave can significantly change the turbulence dynamics.

Figure~\ref{fig:resolvent_structure_3d_velocity}(\sublabel{c}) shows the optimal forcing mode on the same cross-plane as the velocity field in figure~\ref{fig:resolvent_structure_3d_velocity}(\sublabel{a}). The corresponding plane-averaged squared forcing components is plotted in figure~\ref{fig:resolvent_structure_3d_velocity}(\sublabel{d}). Similar to the streamwise-invariant mode discussed above, the optimal forcing for the three-dimensional structure is primarily in the streamwise direction. That is, alternating streamwise accelerations and decelerations induced by nonlinear interactions propagate through the flow and, via the linear amplification mechanism, drive the formation of inclined vortex structures shown in figure~\ref{fig:resolvent_structure_3d}.

Both the two-dimensional and three-dimensional response modes presented above exhibit characteristic features that are consistent with the large-scale roll cells or smaller-scale vortices observed in Langmuir turbulence. For both selected modes, the leading singular value is noticeably larger than the second mode, with $\sigma_1/\sigma_2 \approx 3.4$ for the two-dimensional mode and $\sigma_1/\sigma_2 \approx 2.4$ for the three-dimensional mode. Therefore, at these scales, the optimal mode and its associated amplification dynamics are more pronounced than those of the subsequent modes. The singular value behaviours and the associated low-rank property of the flow are further discussed in \S\,\ref{sec:energy}. For completeness, the secondary mode is shown in Appendix \ref{app:secondary}. 

These results demonstrate that the linearised model successfully captures the dominant coherent structures arising from turbulence--wave interactions. We note again that the physical meaning of the resolvent analyses is different from that of traditional stability analyses or temporal energy growth problems, which describe the process that Langmuir circulations arise from the evolution of small-amplitude, unstable initial perturbations. The results of the resolvent analyses indicate that the Langmuir vortical structures can also be sustained by the linear amplification of harmonic forcing, which may originate from the nonlinearity inherent in the turbulent flow. We also note that, according to the maximum energy amplification (figure~\ref{fig:resolvent_Hnorm_2d}), the three-dimensional Langmuir vortices near the water surface (figure~\ref{fig:resolvent_structure_3d}) are highly sensitive to nonlinear forcing, which indicates their amplification potential through the input--output amplification mechanism shown in figure~\ref{fig:resolvent_structure_3d_velocity} and supports the expectation that inclined, smaller-scale vortical structures are a prominent feature of Langmuir turbulence. Building on these results, we next show that the resolvent model can reproduce certain statistical properties of coherent turbulent structures.

\subsection{Energy distribution across various wavelengths\label{sec:energy}}
In this section, we continue analysing the resolvent model for linearised turbulence--wave interaction and use it as a reduced-order model that retains only the principal forcing and response modes. We demonstrate that, even in this simplified form, the model can reproduce features of turbulence statistics in Langmuir turbulence. Specifically, we focus on vertical velocity fluctuations, because strong vertical mixing is one of the most prominent characteristics of Langmuir turbulence and the vertical turbulence velocity fluctuation intensity is an important metric for quantifying mixing in a wave-driven ocean boundary layer~\citep[see e.g.][]{dasaro2014}.

Here, we consider the system's response to unit-amplitude forcing across all possible wavenumber--frequency triplets, $(k_x, k_z, \omega)$. Although the forcing in a realistic flow requires a weighted sum of input modes, the broadband harmonic forcing can still offer insights into the flow dynamics~\citep{moarref_modelbased_2013}.  By the definition of the Schmidt decomposition~\eqref{eq:svd_T}, the fraction of energy contributed by the $j$-th response mode $\boldsymbol{\psi}_j$ is given by $\sigma_j^2 / (\sum_{k} \sigma^2_k)$. If a few leading modes dominate the energy content of the velocity response, the transfer operator $\mat{T}$ exhibits a low-rank structure, suggesting that the input--output dynamics of the system may be governed by a small number of resolvent modes.

\begin{figure}
  \centering
  \includegraphics{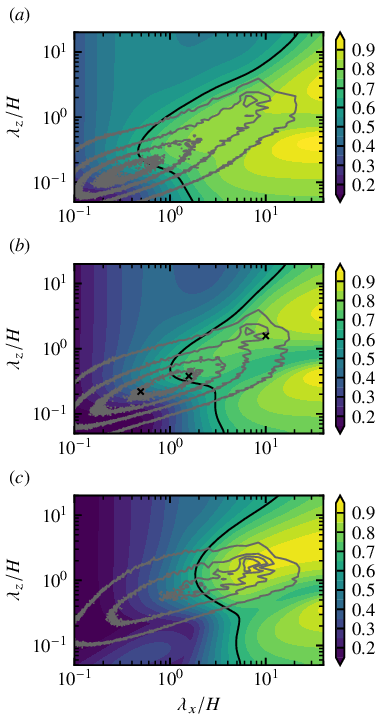}
  \includegraphics{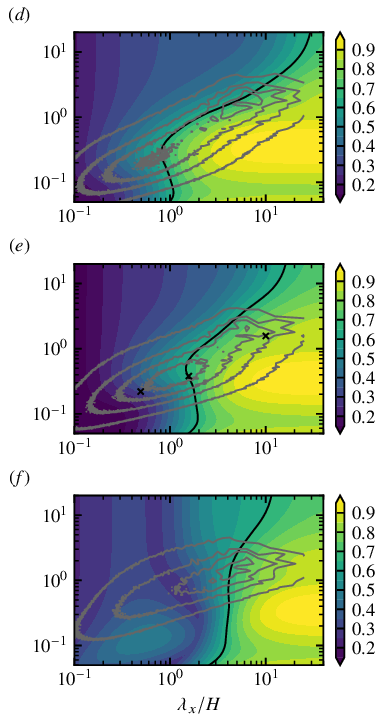}
  \caption{\label{fig:resolvent_svals_ratio}Contours of the energy ratio of the leading response mode relative to the total response, $\sigma^2_1/\sum_j \sigma^2_j$, for different streamwise and spanwise wavelengths at selected phase speeds $c$: (\sublabel{a},\sublabel{d}) $c=U^L(y=-0.05H)$, (\sublabel{b},\sublabel{e}) $c=U^L(y=-0.12H)$ and (\sublabel{c},\sublabel{f}) $c=U^L(y=-0.4H)$. The black line indicates the contour of the $65\%$ energy ratio. The grey contour lines represent the pre-multiplied turbulent kinetic energy spectrum from the LES at the corresponding depths, plotted from $20\%$ to $80\%$ of the maximum values in increments of $20\%$. The two columns correspond to the results for (\sublabel{a}--\sublabel{c}) the case with $\Lat=0.2$ and (\sublabel{d}--\sublabel{f}) the case with $\Lat=0.3$. In (\sublabel{b}) and (\sublabel{e}), the black crosses mark the wavelengths selected for the spectrum of the singular values shown in figure sequences~\ref{fig:resolvent_svals_spectrum}(\sublabel{a}--\sublabel{c}) and \ref{fig:resolvent_svals_spectrum}(\sublabel{d}--\sublabel{f}), respectively.}
\end{figure}

Figure~\ref{fig:resolvent_svals_ratio} shows the energy ratio of the leading mode, $\sigma_1^2 / (\sum_{k} \sigma^2_k)$, for different streamwise and spanwise wavelengths. In this figure, we examine three values of the phase speed $c$, corresponding to the Lagrangian mean velocity $U^L$ at three depths representative of the near-surface and bulk regions of the flow. At all the selected depths, the energy ratio of the leading mode is large over a broad range of streamwise and spanwise wavelengths, which is evident from the large regions where the leading mode accounts for $65\%$ of the total energy (enclosed by the black contours in figure~\ref{fig:resolvent_svals_ratio}). This result demonstrates the broad applicability of a low-rank approximation for various turbulence scales.

Notably, the most energetic scales in LES tend to coincide with the wavenumbers for which the resolvent system is low-rank. To illustrate this point, in figure~\ref{fig:resolvent_svals_ratio}, the pre-multiplied energy density spectrum computed from LES (indicated by the grey contours) is overlaid on the energy ratio of the leading resolvent mode. As seen in the figure sequences~\ref{fig:resolvent_svals_ratio}(\sublabel{a}--\sublabel{c}) and~\ref{fig:resolvent_svals_ratio}(\sublabel{d}--\sublabel{f}), the wavelength region where the system is strongly low-rank shifts towards larger-scale motions with increasing depth. This trend is consistent with the LES results, which show that the turbulence structures grow in size with increasing depth and that larger coherent structures become more dominant at greater depths. For example, as shown in figure~\ref{fig:resolvent_svals_ratio}(\sublabel{a}), near the surface ($y=-0.05H$), the leading mode accounts for approximately $60\%$ of the total energy at the energy spectrum peak of LES ($\lambda_x=0.4H$, $\lambda_z=0.18H$). At deeper depths (e.g.\ $y=-0.4H$, figure~\ref{fig:resolvent_svals_ratio}\sublabel{c}), short wavelengths no longer exhibit low-rank behaviour. However, the system remains highly low-rank at the energy peak of LES ($\lambda_x= 7H$, $\lambda_z=H$), corresponding to large-scale circulation cells. This overlap of the energetic turbulence scales with the low-rank operators, also observed in turbulent channel flows~\citep{moarref_modelbased_2013}, suggests that the key dynamics in Langmuir turbulence may be effectively captured by the leading resolvent mode.

\begin{figure}
  \centering
  \includegraphics{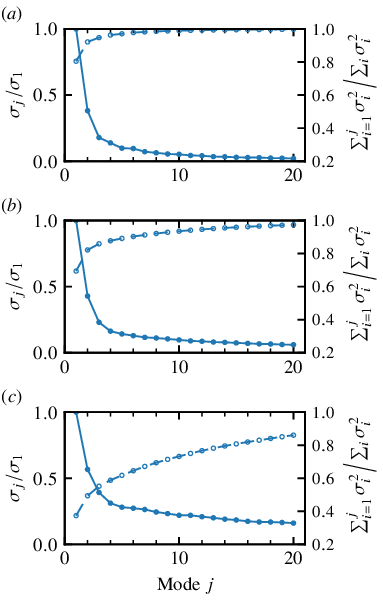}
  \includegraphics{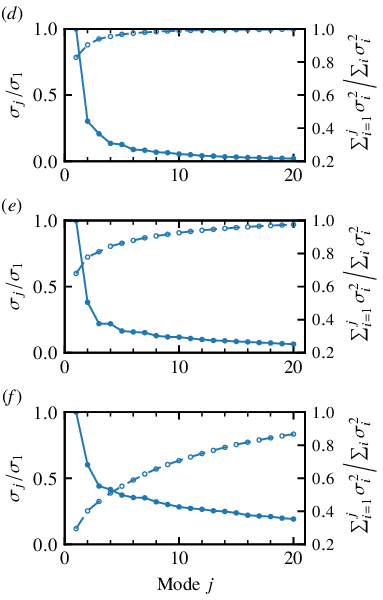}
  \caption{\label{fig:resolvent_svals_spectrum}The normalised singular values, $\sigma_j/\sigma_1$, and the cumulative energy ratio, $\sum_{i}^{j} \sigma_i^2/\sum_i \sigma_i^2$, of the first twenty resolvent modes at selected streamwise and spanwise wavelengths: (\sublabel{a},\sublabel{d}) $(\lambda_x, \lambda_z) = (10H, \pi H/2)$, (\sublabel{b},\sublabel{e}) $(\lambda_x, \lambda_z) = (\pi H/2, 4\pi H/33)$ and (\sublabel{c},\sublabel{f}) $(\lambda_x, \lambda_z) = (0.5H, 0.2H)$. The cumulative energy ratio is plotted against the axis on the right. The two columns correspond to the results for (\sublabel{a}--\sublabel{c}) the case with $\Lat=0.2$ and (\sublabel{d}--\sublabel{f}) the case with $\Lat=0.3$. For both cases, the phase speeds are chosen as $c=U^L(y=-0.12H)$. The three modes are marked by black crosses in figures~\ref{fig:resolvent_svals_ratio}(\sublabel{b}) and~\ref{fig:resolvent_svals_ratio}(\sublabel{e}) for cases $\Lat=0.2$ and $\Lat=0.3$, respectively.}
\end{figure}

To further examine the low-rank characteristics of the system, the normalised singular values and the cumulative energy ratio of the leading twenty modes are plotted in figure~\ref{fig:resolvent_svals_spectrum} for selected wavenumber--frequency triplets (marked by black crosses in figures~\ref{fig:resolvent_svals_ratio}\sublabel{b} and \ref{fig:resolvent_svals_ratio}\sublabel{e}). For large-scale motions, e.g.\ those shown in figures~\ref{fig:resolvent_svals_spectrum}(\sublabel{a},\sublabel{b}) and \ref{fig:resolvent_svals_spectrum}(\sublabel{d},\sublabel{e}), the leading singular value is appreciably larger than the second and third modes. The remaining singular values decay rapidly towards zero. Correspondingly, the energy content is well captured by the first few modes, with the leading mode contributing a significant fraction. For small-scale motions, as shown in figure~\ref{fig:resolvent_svals_spectrum}(\sublabel{c},\sublabel{f}), although the largest singular value is still approximately twice as large as the second mode, the decay of singular values of the higher modes is more gradual. This behaviour indicates that the low-rank nature of the resolvent system is more pronounced at larger scales, whereas fully capturing small-scale dynamics may require retaining a greater number of modes. Nevertheless, as discussed above, the low-rank approximation can be effective in revealing key characteristics of Langmuir turbulence given that the energetic scales tend to be low-rank.

As shown in \S{}\,\ref{sec:modes}, the resolvent model-predicted coherent structures exhibit strong vertical velocity intensity, suggesting that the model may offer information about the vertical velocity component of turbulence fluctuations.
Considering that the system exhibits a low-rank property across a range of scales, when only the principal mode is retained, the pre-multiplied energy density of the response to the broadband forcing is given by~\citep{moarref_modelbased_2013}:
\begin{equation}
  E_{v}(y; k_x, k_z) = \int k_x k_z {\left(\sigma_1 \left| v_1(y; k_x, k_z, \omega) \right| \right)}^2\,\mathrm{d}\omega,
\end{equation}
where $v_1$ denotes the vertical component of the principal response velocity $\boldsymbol{\phi}_1$. 
The function $E_v(y; k_x, k_z)$ is a two-dimensional energy spectrum at a vertical location $y$, indicating the energy content per logarithmic interval of streamwise and spanwise wavenumbers. This integral can be reformulated with respect to the phase speed $c$ using the relation $\omega = c k_x$, yielding
\begin{equation}
  E_{v}(y; k_x, k_z) = \int k_x^2 k_z {\left(\sigma_1 \left| v_1(y) \right| \right)}^2 \, \mathrm{d}c. \label{eq:Evv_kxkz}
\end{equation}
The integrand in~\eqref{eq:Evv_kxkz} is consistent with the pre-multiplied energy density formulation (2.13) in~\citet{moarref_modelbased_2013}. A further integration provides the pre-multiplied energy density spectrum with respect to the spanwise wavenumber:
\begin{equation}
  E_{v}(y; k_z) = \iint k_x^2 k_z {\left(\sigma_1 \left| v_1(y) \right| \right)}^2 \, \mathrm{d}\log(k_x)\mathrm{d}c. \label{eq:Evv_kz}
\end{equation}

\begin{figure}
  \centering
  \includegraphics{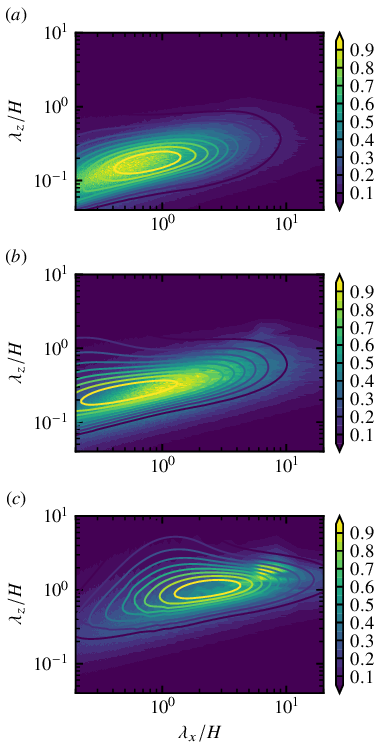}
  \includegraphics{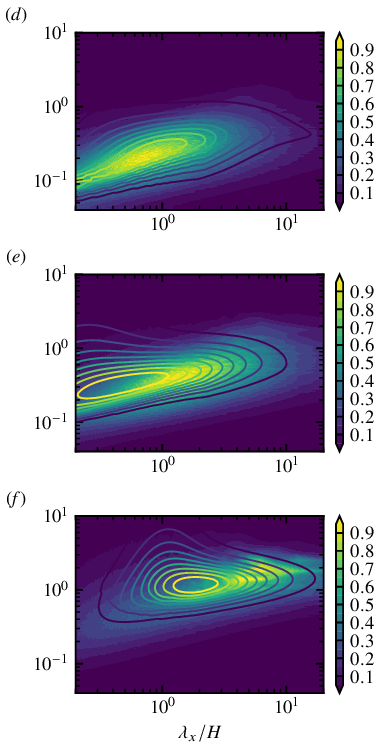}
  \caption{\label{fig:energy_spectra_2d}Comparisons of the normalised pre-multiplied energy spectrum of the vertical velocity $v$ with respect to the streamwise and spanwise wavelengths, $\lambda_x$ and $\lambda_z$, between the model prediction (contours, from \eqref{eq:Evv_kxkz}) and LES (colour). The results are shown for cases (\sublabel{a}--\sublabel{c}) $\Lat=0.2$ and (\sublabel{d}--\sublabel{f}) $\Lat=0.3$ at selected depths: (\sublabel{a},\sublabel{d}) $y=-0.05H$, (\sublabel{b},\sublabel{e}) $y=-0.12H$ and (\sublabel{c},\sublabel{f}) $y=-0.4H$. Each spectrum is normalised by its respective maximum value and the contour lines of the model-predicted spectrum are drawn from $0.1$ to $0.9$ in increments of $0.1$.}
\end{figure}

Figure~\ref{fig:energy_spectra_2d} compares the pre-multiplied two-dimensional energy spectrum of $v$ obtained from LES with the energy distribution predicted by the resolvent-based model~\eqref{eq:Evv_kxkz}. As shown by the energy spectrum evaluated from LES, the turbulence structures span a wide range of streamwise wavelengths but are confined to a narrower range of spanwise scales, which manifests the streamwise elongation of vortices in Langmuir turbulence. For all the depths and turbulent Langmuir numbers considered, the energy distribution is centred along a straight line on the log-log scale, indicating that the ratio of the streamwise to spanwise wavelengths approximately follows a power-law relationship. For the energy distribution predicted by the model, the slope of the contours closely matches that of the simulation, indicating that the model captures this scaling trend reasonably well. However, we do observe a slight discrepancy in the peak streamwise wavelength. For example, the model tends to underestimate the streamwise wavelengths of turbulent structures at deeper depths, as shown in figure~\ref{fig:energy_spectra_2d}(\sublabel{c},\sublabel{f}). In addition, although the model captures the trend of the increased length scales of turbulence structures at greater depths, it fails to capture the energy peak associated with the large-scale Langmuir cells ($\lambda_z\approx 2H$). This discrepancy may be due to the current modelling assumptions, particularly the use of unit-amplitude broadband forcing for all scales, which may underrepresent the contribution from very large-scale motions and skew the results towards near-surface, smaller-scale turbulence structures.

\begin{figure}
  \centering
  \includegraphics{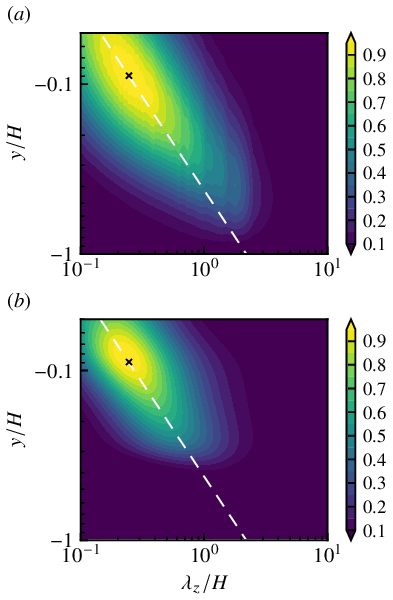}
  \includegraphics{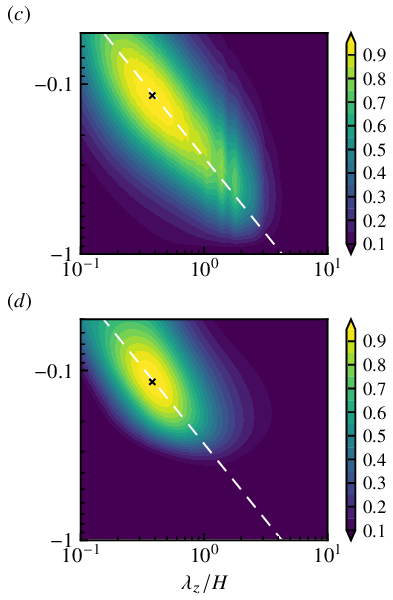}
  \caption{\label{fig:energy_spectra_y_1d}Comparisons of the normalised one-dimensional pre-multiplied energy spectrum of the vertical velocity $v$ with respect to the spanwise wavelength $\lambda_z$ between (\sublabel{a},\sublabel{c}) LES and (\sublabel{b},\sublabel{d}) model prediction from~\eqref{eq:Evv_kz}, for (\sublabel{a},\sublabel{b}) $\Lat=0.2$ and (\sublabel{c},\sublabel{d}) $\Lat=0.3$. Each spectrum is normalised by its respective maximum value. The white dashed lines indicate the fitted power-law relationships between the peak spanwise wavelength and the vertical coordinate $y$ for the LES data: (\sublabel{a}) $y=0.38(\lambda^{peak}_z)^{1.1}$ for $\Lat=0.3$, and (\sublabel{c}) $y=0.27(\lambda^{peak}_z)^{0.9}$ for $\Lat=0.3$. The black crosses in (\sublabel{a}) and (\sublabel{c}) mark the ($\lambda_z$, $y$) where the energy density peaks in LES for cases $\Lat=0.2$ and $\Lat=0.3$, respectively. For the purposes of comparison, the white dashed lines and the black crosses, indicating the characteristic features of the LES energy spectra, are overlaid on the corresponding model-predicted spectra in (\sublabel{b}) and (\sublabel{d}).}
\end{figure}

To further examine the model's capability, we assess its performance in capturing the variations in dominant spanwise length scales with depth, which are closely related to the cross-wind spacings of circulating structures. Figures~\ref{fig:energy_spectra_y_1d}(\sublabel{a}) and~\ref{fig:energy_spectra_y_1d}(\sublabel{c}) show the energy density spectrum with respect to the spanwise wavelength $\lambda_z$ from LES for cases $\Lat=0.2$ and $\Lat=0.3$, respectively. With increasing depth, the spanwise length scales of the vertical velocity $v$ increase, which is consistent with previous findings of the shift of dominant structures towards larger-scale motions at greater depths. The energy density of $v$ peaks near the surface at $y \approx -0.1H$, further confirming that most of the vertical kinetic energy in Langmuir turbulence is concentrated in the near-surface region and is associated with smaller-scale turbulence vortices. While large-scale Langmuir cells are still important in the bulk region for mixing and mixed-layer deepening, they contribute less to the total vertical kinetic energy than do smaller-scale vortices. This result underscores the importance of modelling smaller-scale turbulence structures near the surface. 

The model-predicted energy distributions, computed using~\eqref{eq:Evv_kz}, are shown in figures~\ref{fig:energy_spectra_y_1d}(\sublabel{b}) and~\ref{fig:energy_spectra_y_1d}(\sublabel{d}) for cases $\Lat=0.2$ and $\Lat=0.3$, respectively. We find that, for both cases, the model-predicted energy distributions (figure~\ref{fig:energy_spectra_y_1d}\sublabel{b},\sublabel{d}) are remarkably similar to the peak energy scales in the LES (figure~\ref{fig:energy_spectra_y_1d}\sublabel{a},\sublabel{c}), particularly in the near-surface region where the vertical turbulence density is strong. First, the model-predicted spanwise wavelength and the depth at which the energy density reaches the maximum agree reasonably well with the LES data. Moreover, the model captures the depth-dependent variations in the dominant spanwise scales. In figures~\ref{fig:energy_spectra_y_1d}(\sublabel{a},\sublabel{b}) and~\ref{fig:energy_spectra_y_1d}(\sublabel{c},\sublabel{d}), we plot the power-law fits of the peak spanwise wavelengths in LES for cases $\Lat=0.2$ and $\Lat=0.3$, respectively. The contours of the model-predicted energy density, as shown in figure~\ref{fig:energy_spectra_y_1d}(\sublabel{b},\sublabel{d}), follow the same trends indicated by the curves from the simulation data. These results indicate that the resolvent model effectively predicts the dominant spanwise length scales of vertical velocity fluctuations, particularly in the near-surface region. In deeper regions (e.g.~$y<-0.4H$), where the LES shows the presence of large-scale Langmuir cells with $\lambda_z>H$ (figure~\ref{fig:energy_spectra_y_1d}\sublabel{a},\sublabel{c}), the model does not fully capture the energy contributions from these structures (figure~\ref{fig:energy_spectra_y_1d}\sublabel{b},\sublabel{d}). While this difference reflects the model's bias towards near-surface structures in predicting the spanwise energy distribution, similar to the two-dimensional spectra discussed above, the model remains valuable for understanding the energetically important motions in Langmuir turbulence, as it accurately captures the dominant spanwise scales associated with vortical structures near the surface, where the vertical turbulence intensity is strong.

To summarise, in this section, we have presented results on the low-rank behaviour of the linearised system and applied the low-rank property to investigate the spectral response of the system under broadband harmonic forcing. Despite the simplified assumptions regarding the harmonic forcings representing the nonlinear effects, the model is effective in capturing key features of the length scales of the dominant velocity component, i.e.\ the vertical velocity. These results indicate that the linearised model effectively reproduces the key dynamics underlying turbulence--wave interactions, supporting the interpretation that near-surface quasi-streamwise vortices in Langmuir turbulence can arise from the linear amplification of nonlinear forcing. Additionally, the findings demonstrate the applicability of the low-rank approximation and the predictive capability of the resolvent model for Langmuir turbulence.



\section{Conclusions and discussion\label{sec:conclusions}}
In this paper, we present a resolvent-based analysis of Langmuir turbulence, a wave-forced turbulent boundary layer in upper oceans, using linearised perturbation equations derived from the CL equations. The linearised model includes the force of a prescribed surface wave and incorporates the profiles of the mean velocity of the turbulent shear current and eddy viscosity obtained from LES, enabling a more realistic representation of the time-averaged state of Langmuir turbulence than existing modal analyses of Langmuir circulations. By employing the resolvent analysis, we obtain the linear amplification behaviour of the system's input--output dynamics, which may be considered the excitation of time-invariant, wave-like propagating coherent structures by turbulence nonlinearity. To our knowledge, this work extends the application of resolvent analysis to the turbulence--wave interaction problem for the first time.

The analysis provides new insights into the sustained structure and energetics of Langmuir turbulence, complementing the instability-based interpretation of Langmuir turbulence.
By analysing the maximum amplification of the resolvent model, we find that the amplified structures can be categorised into two regimes according to their spanwise length scales. The structures with very large spanwise wavelengths ($\lambda_z > H$) achieve maximum amplification for streamwise-invariant motions that extend nearly the full boundary layer depth. These motions correspond to large-scale, quasi-two-dimensional Langmuir cells with strong spanwise and vertical motions. The associated optimal forcing mode indicates that these rolls are driven by forcings in the streamwise direction. The amplifications at smaller spanwise wavelengths ($\lambda_z < H$) are strong over a range of finite streamwise wavelengths, indicating that various three-dimensional coherent structures can be responsive to the amplification mechanism. These three-dimensional coherent structures take the form of quasi-streamwise vortices inclined upwards in the downwind direction and are similarly driven primarily by streamwise forcings. The optimal response modes are similar to the characteristic vortical structures in real flows, suggesting that the input--output amplification mechanism captured by the resolvent model can represent the coherent structure dynamics in Langmuir turbulence.

The dynamics revealed by the resolvent analysis, i.e.\ the excitation of coherent structures by nonlinear forcing, offer a more general interpretation of Langmuir turbulence, and, more broadly, turbulence--wave interactions, than does the classic CL-II mechanism. In addition to capturing the large-scale rolls that have been the primary focus of existing stability analyses, the resolvent analysis also elucidates the mechanisms underlying the smaller-scale vortical structures that are prominent near the surface. Furthermore, unlike linear stability or transient energy analyses, which describe Langmuir circulations as the result of growing initial perturbations, resolvent analyses demonstrate that similar structures can form through linear amplification of persistent harmonic forcing and that such forcing may be due to nonlinear interactions in turbulence. This forced response mechanism can be particularly relevant to the fully developed Langmuir turbulence, where sustained forcing continually excites coherent structures.  As such, these findings complement the classic CL-II mechanism by providing a more comprehensive view of the multiscale coherent structures in Langmuir turbulence. 

Building on the finding that coherent structures can be forced by sustained nonlinear forcing, we consider the system's response to broadband harmonic forcing. When the principal response modes are integrated across all wavelengths and frequencies, the superimposed response is found to reproduce the key features of the vertical velocity spectrum observed in LES.\@ This result is possible because the resolvent operator is low-rank over the wavenumbers where turbulence fluctuations are most energetic, allowing the dominant dynamics to be captured by the principal modes.

Specifically, by examining the spectra predicted by the superimposed response, we find that the model-predicted two-dimensional spectra with respect to the streamwise and spanwise wavelengths are in qualitative agreement with the real spectra from LES.\@ Moreover, the model accurately captures the variations in the dominant spanwise length scales of vertical velocity fluctuations with depth, particularly in the near-surface region where the vertical kinetic energy is concentrated. These results suggest that the dominant dynamics determining the structures of the vertical velocity fluctuations in Langmuir turbulence are likely governed by linear amplification mechanisms. We also note that, although the simplified assumption of unit-amplitude broadband forcing successfully enables valuable predictions of the multiscale features of Langmuir turbulence, the model still shows discrepancies in the streamwise length scales and does not fully capture the energy of large-scale Langmuir cells in the bulk region. These limitations prompt further improvement in input forcing modelling, which should be weighted to reflect the nonlinear interactions in turbulence more realistically. One promising direction is to use a data-driven method to determine the appropriate forcing or eddy viscosity terms~\citep{zare_colour_2017,gupta_linearmodelbased_2021}. Such approaches can potentially enhance the predicted power of resolvent-based models for complex turbulence--wave interaction systems.

The results presented in this study demonstrate that a resolvent model using the turbulent mean flow and eddy viscosity as the base state is a promising approach for understanding turbulence--wave interaction dynamics. Both the coherent structures and the statistical features of the energy distribution predicted by the model indicate that the linear amplification of nonlinear forcing is a physically meaningful formation mechanism of Langmuir vortices. This resolvent modelling framework shows the feasibility of developing efficient tools to predict turbulence statistics in the ocean surface boundary layer under the influence of surface gravity waves. In this framework, we construct a model based on the CL equations, which use a wave-phase-averaged approach to quantify the wave effect without explicitly resolving the wave orbital motions. However, a crucial feature of turbulence--wave interactions is the strain rate associated with wave orbital motion, which induces turbulence fluctuations with time scales comparable to the wave period and results in wave-phase-dependent variations in turbulent coherent structures~\citep{xuan2019a,xuan2020,smeltzer2023,xuan_effect_2024}. To capture these effects, extending the current linearised system into a wave-phase-resolved framework is desirable. Such an extension may be achieved either by formulating the governing equations in a curvilinear coordinate system~\citep{teixeira2002,cao_numerical_2021} or by incorporating the wave boundary conditions through appropriate wave-phase-dependent perturbation expansions~\citep{fedele_momentary_2022,xuan_effect_2024}. Additionally, the modulation of the wind-induced surface shear by wave geometry~\citep{thais1996,yang2013} may be incorporated as a harmonic forcing to represent coupled air--sea--wave interactions. These future developments would enable resolvent-based models to resolve wave-phase-dependent turbulence structures and to capture the turbulence--wave interaction processes more completely.

\begin{bmhead}[Declaration of interests]
The authors report no conflict of interest.
\end{bmhead}

\begin{appen}
\section{Alternative formulation of the perturbation equation}\label{app:alternative_formulation}
Equation~\eqref{eq:perturbation_eq} can be reformulated to explicitly include a vortex force term. Applying the vector identity $\bnabla(\boldsymbol{u}\bcdot \boldsymbol{U}^s) = (\boldsymbol{u} \bcdot \bnabla) \boldsymbol{U}^s + (\boldsymbol{U}^s \bcdot \bnabla) \boldsymbol{u} + \boldsymbol{u} \times (\bnabla \times \boldsymbol{U}^s) + \boldsymbol{U}^s \times (\bnabla \times \boldsymbol{u})$, the perturbation equation~\eqref{eq:perturbation_eq} becomes
\begin{equation}
  \frac{\partial \boldsymbol{u}}{\partial t} + (\boldsymbol{U} \bcdot \bnabla) \boldsymbol{u} 
  + (\boldsymbol{u} \bcdot \bnabla) \boldsymbol{U} 
   = - \nabla \tilde{p}
  + \nu \bnabla \bcdot \left[ \frac{\nu_T}{\nu} (\bnabla \boldsymbol{u} + \bnabla\boldsymbol{u}^\mathrm{T}) \right] + \boldsymbol{U}^s \times (\bnabla \times \boldsymbol{u})
  + \boldsymbol{d}, \label{eq:alt_perturbation_eq}
\end{equation}
where the modified pressure $\tilde{p}$ absorbs the gradient term $\bnabla(\boldsymbol{u}\bcdot \boldsymbol{U}^s)$ and $\boldsymbol{U}^s \times (\bnabla \times \boldsymbol{u})$ is the vortex force acting on the perturbations. It is also straightforward to derive the linearised equation in the above form directly from~\eqref{eq:CL_LES}, excluding the SGS term. The vortex force in the above perturbation equation, $\boldsymbol{U}^s \times (\bnabla \times \boldsymbol{u})$, corresponds to the perturbation component of total vortex force $\boldsymbol{U}^s \times (\bnabla \times \boldsymbol{v})$ in~\eqref{eq:CL_LES}.

\section{Rectangular spectral collocation method for discretising the linearised system}\label{app:collocation}
The rectangular spectral collocation method~\citep{driscoll_rectangular_2016}, a generalised approach to incorporate boundary conditions into the spectral collocation discretisation of differential equations, is employed to discretise the linearised perturbation equations in the present study. Consider a linear differential equation of order $m$ for a function $u(x)$:
\begin{equation}
  a_0(x) u(x) + a_1(x) u'(x) + \dots + a_m (x) u^{(m)}(x) = f(x), \label{eq:collocation_example}
\end{equation}
where $a_i(x)$ ($i=0\dots m$) are the coefficients of each derivative order and $f(x)$ on the right-hand side is a source term. 
This equation requires $m$ boundary conditions, specified as
\begin{equation}
  l_i(u, u', \dots, u^{(m-1)})=g_i, \quad i=1,\dots,m,
\end{equation}
where $l_i$ denote the linear functionals defining the boundary constraints and $g_i$ are prescribed scalars.

The function $u$ is discretised at $N + m$ Chebyshev points of the second kind $\{x_j\}$, given by
\begin{equation}
  x_j = -\cos\left(\frac{j\pi}{N + m -1}\right), \quad j=0,\dots,N + m -1. \label{eq:collocation_nodes_second}
\end{equation}
The spectral collocation method evaluates the differential equation on a different set of nodes, which are $N$ Chebyshev nodes of first kind $\{\breve{x}_j\}$, defined by
\begin{equation}
  \breve{x}_j = -\cos \left(\frac{(j+\frac{1}{2})\pi}{N} \right), j=0, \dots, N - 1. \label{eq:collocation_nodes_first}
\end{equation}
The conversion between the two sets of collocation grids is obtained using the rectangular differential operator of the form $\mat{P}\mat{D}^k$, where $\mat{P}$ is a barycentric resampling matrix that interpolates the polynomial represented by the nodal values at $\{x_j\}$ to $\{\breve{x}_j\}$ (see \citealt{berrut_barycentric_2004} and \citealt{driscoll_rectangular_2016} for the formulation), and $\mat{D}^k$ is the $k$-th order differentiation matrix evaluated at~\eqref{eq:collocation_nodes_second}~\citep{trefethen_spectral_2000}. The operator $\mat{P}\mat{D}^k$, which downsamples the derivatives from $N+m$ points to $N$ nodes, has dimensions $N \times (N+m)$. After adding up all the differential terms, the $m$ boundary conditions are appended to the system to form a rectangular system matrix. For example, a Dirichlet boundary condition, $u=0$, at the first node is enforced by appending
\begin{equation}
  [1 \quad 0 \quad \cdots \quad 0]\begin{bmatrix}
    u_0 \\
    u_1 \\
    \vdots \\
    u_{N-1}
  \end{bmatrix} = 0,
\end{equation}
where $[u_0, u_1, \dots, u_{N-1}]^\mathrm{T}$ denotes the discretised solution vector. Note that the Chebyshev points of the second kind  $\{x_j\}$~\eqref{eq:collocation_nodes_second} include the domain boundaries. Neumann conditions can be similarly enforced by appending the corresponding rows (first and last low) of the differentiation matrix $\mat{D}^{k}$.
The resulting system takes the following form:
\begin{equation}
  \left[\begin{array}{c}
    P(a_0\mat{I} + a_1\mat{D}^1 + \dots + a_m\mat{D}^m) \\
    \mat{L}
  \end{array}\right] \boldsymbol{u} = 
  \left[\begin{array}{c}
    P\boldsymbol{f} \\
    \boldsymbol{g}
  \end{array}\right],
\end{equation}
where $\mat{L}$ represents the discretised boundary constraint functionals and $\boldsymbol{g}$ contains the boundary values.

In the present study, when discretising the linearised perturbation equations \eqref{eq:perturbation_eq}, $v$ and $\omega_y$ are discretised on $N+4$ and $N+2$ Chebyshev points of the second kind, respectively, as the equations for $v$ and $\omega_y$ are fourth-order and second-order in $y$, respectively. The rectangular collocation method resamples the function values and enforces the equations on $N$ nodes. 
It should be noted that the downsampling does not lead to the loss of information. A function represented by $N + m$ nodal values corresponds to a polynomial of degree at most $N + m -1$. For a differential equation of order $m$, the associated polynomial can only be enforced at $N$ degrees of freedom. The remaining $m$ degrees of freedom are constrained by the boundary conditions. Therefore, the downsampling is a natural approach for enforcing the differential equations in the domain interior while preserving the required degrees of freedom for boundary conditions. Further discussions about the resampling and the relation of the rectangular collocation method to other approaches, such as the Chebyshev tau method, are available in~\citet{driscoll_rectangular_2016}.

\section{Secondary resolvent response mode and forcing mode}\label{app:secondary}
In this section, we present the secondary mode for selected wavenumber--frequency triplets, complementing the principal mode discussed in \S\,\ref{sec:modes}. Although the principal mode has the highest amplification and is more prominent in the flow, the subsequent modes also contribute to the overall flow structure. 

\begin{figure}
  \centering
  \includegraphics{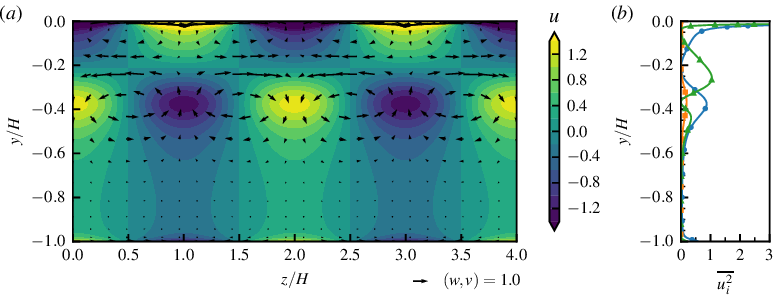}
  \includegraphics{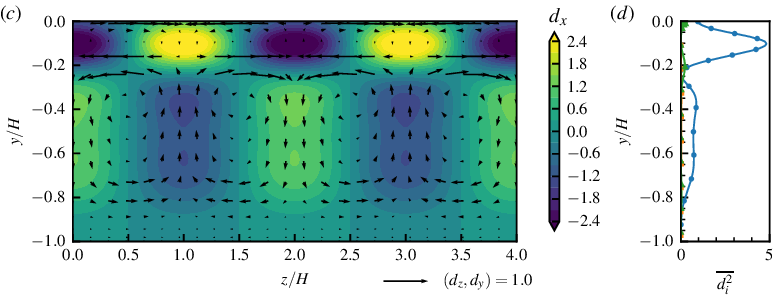}
  \caption{\label{fig:resolvent_secondary_structure_2d}Structures of the secondary (\sublabel{a}) velocity response $\boldsymbol{u}=(u,v,w)$ and (\sublabel{c}) input forcing $\boldsymbol{d}=(d_x,d_y,d_z)$ for case $\Lat=0.2$ at $(k_x, k_z, \omega)=(0, 2\upi/H, 0)$. In (\sublabel{a}) and (\sublabel{c}), the contours represent the streamwise component, $u$ and $d_x$, respectively; the vectors represent the cross-stream components, $(w,v)$ and $(d_z, d_y)$, respectively. The vertical variations in the mean squared response velocity and forcing components are plotted in (\sublabel{b}) and (\sublabel{d}), respectively: streamwise component ($\overline{u^2}$ or $\overline{d_x^2}$, \fullcirc), vertical component ($\overline{v^2}$ and $\overline{d_y^2}$, \fullsqr) and spanwise component ($\overline{w^2}$ and $\overline{d_z^2}$, \fulltri).}
\end{figure}

Figure~\ref{fig:resolvent_secondary_structure_2d} shows the secondary response and forcing structures for the two-dimensional, streamwise-invariant mode examined in \S\,\ref{sec:modes}, whose principal response and forcing are presented in figure~\ref{fig:resolvent_structure_2d}. The flow response is dominated by the spanwise and streamwise velocity components, exhibiting as two vertically stacked layers of vortical motions with accompanying variations in the streamwise velocity. The vortices near the surface and deeper in the domain are aligned in the vertical direction and rotate in opposite directions. Near the surface, the streamwise velocity is positive in the converging zone and negative in the diverging zone. In the lower layer, positive streamwise velocity is found to accompany the downward motion between counter-rotating vortices, which transports high-momentum fluid downward, whereas negative streamwise velocity corresponds to upward motions. This flow response corresponds to nonlinear forcing that is dominated by the streamwise component, i.e.\ alternating accelerations and decelerations in the streamwise momentum equation. The forcing also exhibits a two-layer structure. 
The two-layer structure in the secondary mode is also observed in conventional turbulent boundary layers without the wave forcing~\citep[see e.g.][]{mckeon_criticallayer_2010,abreu_resolvent_2020}. 

\begin{figure}
  \centering
  \includegraphics{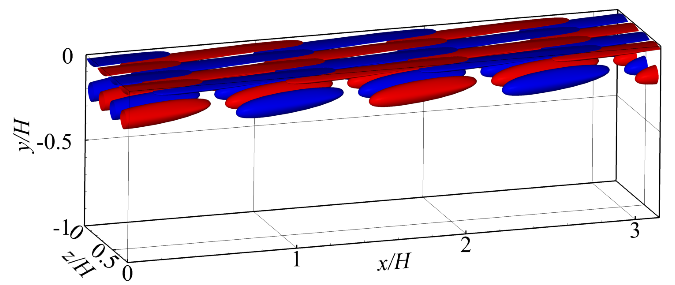}
  \caption{\label{fig:resolvent_secondary_structure_3d}Vortex structures of the secondary response mode for case $\Lat=0.2$ with $k_x H=4$, $k_z H = 16.5$ and $\omega=k_x U^L(y=-0.12H)=44.9 u_*/H$. The vortex structures are elucidated using the iso-surfaces of the $Q$-criterion ($10\%$ of the maximum value), with red and blue indicating positive and negative streamwise vorticity, respectively.}
\end{figure}

\begin{figure}
  \centering
  \includegraphics{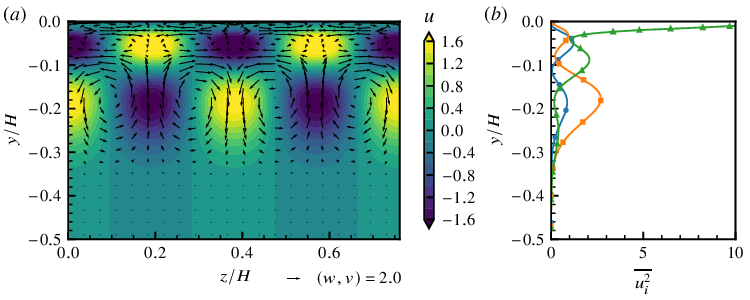}
  \includegraphics{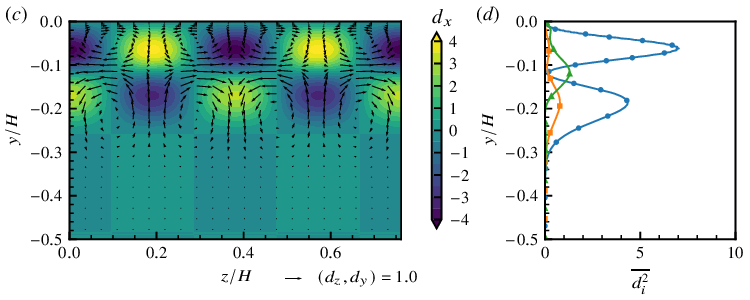}
  \caption{\label{fig:secondary_structure_1}Cross-plane structures of the secondary (\sublabel{a}) response $\boldsymbol{u}=(u, v, w)$ and (\sublabel{c}) input forcing $\boldsymbol{d}=(d_x, d_y, d_z)$ shown in figure~\ref{fig:resolvent_secondary_structure_3d}, plotted at $x=(\upi/12)H$. In (\sublabel{a}) and (\sublabel{c}), the contours represent the streamwise component, $u$ and $d_x$, respectively; the vectors represent the cross-stream components, $(w,v)$ and $(d_z, d_y)$, respectively. The vertical variations in the plane-averaged squared response velocity and forcing components are plotted in (\sublabel{b}) and (\sublabel{d}), respectively: streamwise component ($\overline{u^2}$ or $\overline{d_x^2}$, \fullcirc), vertical component ($\overline{v^2}$ and $\overline{d_y^2}$, \fullsqr) and spanwise component ($\overline{w^2}$ and $\overline{d_z^2}$, \fulltri).}
\end{figure}

Figure~\ref{fig:resolvent_secondary_structure_3d} shows the vortical structures of the secondary three-dimensional response mode for the wavenumber--frequency triplet shown in figure~\ref{fig:resolvent_structure_3d}. This secondary response exhibits as two layers of quasi-streamwise vortices in the near-surface region, with each layer consisting of an array of counter-rotating vortex pairs. From the velocity field on the vertical cross-plane (figure~\ref{fig:secondary_structure_1}\sublabel{a}) and the energy of the three velocity components (figure~\ref{fig:secondary_structure_1}\sublabel{b}), we find that the secondary response is dominated by the spanwise and vertical motions. The streamwise velocity component is relatively weak. The spatial pattern of the streamwise velocity, with positive $u$ in the downwelling regions and negative $u$ in the upwelling zones, aligns with the vertical transport of streamwise momentum induced by the vortical motions. The forcing pattern and the plane-averaged mean squared forcing magnitudes are shown in figures~\ref{fig:secondary_structure_1}(\sublabel{c}) and \ref{fig:secondary_structure_1}(\sublabel{d}), respectively. The streamwise forcing component is dominant, again exhibiting a two-layer distribution. For this scale, the secondary mode shares similar dynamics with the principal mode, i.e.\ quasi-streamwise vortices are driven via the linear amplification of streamwise nonlinear forcing. Additionally, we observe that the dominance of vortical response and streamwise forcing in the secondary mode is applicable to small-scale motions whose wavelengths are smaller than domain depth $H$.
\end{appen}

\bibliographystyle{jfm}
\bibliography{langmuir_resolvent}

\end{document}